\def\noeditingmarks{} 
\def\buildtechreport{}

\documentclass[10pt,twocolumn]{article}
\usepackage{sigcomm-submit}
\usepackage{lastpage}

\usepackage[pdftex]{graphicx}
\usepackage{subfigure}
\usepackage{booktabs}  %
\usepackage{dcolumn}   %
\usepackage{multirow}  %
\usepackage{rotating}
\usepackage{xspace}
\usepackage{hyphenat}  %
\usepackage{color}
\usepackage{enumerate}

\usepackage{textcomp}
\usepackage{amsmath}
\usepackage{amssymb}
\usepackage{amsfonts}
\usepackage{wasysym}
\usepackage{lastpage}
\usepackage{tabularx}
\usepackage{pifont}
\usepackage{microtype} %
\usepackage{mathtools} %
\DeclarePairedDelimiter\ceil{\lceil}{\rceil}
\DeclarePairedDelimiter\floor{\lfloor}{\rfloor}
\usepackage{url}

\newcommand{\yaogllmname}{Yao+\textsc{gllm}\xspace}

\usepackage{ulem}
\usepackage[table]{xcolor}
\usepackage{colortbl} %
\usepackage{array}    %

\usepackage{float}
\usepackage{dblfloatfix}

\usepackage[noend]{algpseudocode}
\algrenewcomment[1]{\hfill// #1}%
\algtext*{Function}
\algrenewcommand{\algorithmicrequire}{\textbf{Input:}}
\algrenewcommand{\algorithmicensure}{\textbf{Output:}}
\algnewcommand{\LineComment}[1]{\State // #1}

\usepackage[noeka]{mathrmletter}
\usepackage{arydshln}

\usepackage{bigdelim}
\usepackage{siunitx}

\newif\ifintrouble
\newif\ifcuttext

\ifintrouble
\cuttexttrue
\else
\cuttextfalse
\fi

\newcommand{\textred}[1]{\begingroup \color{red} #1\endgroup}
\ifx\noeditingmarks\undefined
   \newcommand{\pgwrapper}[2]{\textred{#1: #2}}
   \newcommand{\pgwrapperb}[1]{\textbf{#1}}
\else
   \newcommand{\pgwrapperb}[1]{}
   \newcommand{\pgwrapper}[2]{}
\fi

\ifx\noeditingmarks\undefined

\else

\fi

\ifx\noeditingmarks\undefined

\else

\fi

\newcommand{\sys}{{Pretzel}\xspace}
\newcommand{\Sys}{\sys}

\newcommand{\goodcitationsize}{\fontsize{8.75}{10.5}\selectfont}

\def\hn{\usefont{OT1}{phv}{mc}{n}\selectfont}

\newcommand{\mpfont}{\hn\scriptsize}
\newcommand{\MPworker}[2]{{\color{#1}\vrule\vrule}{\marginpar{\color{#1}\mpfont #2}}}
\ifx\noeditingmarks\undefined
    \newcommand{\MP}[1]{\MPworker{red}{#1}}
    \newcommand{\MPla}[1]{\MPworker{brown}{#1}}
    \newcommand{\MPtg}[1]{\MPworker{blue}{#1}}
    \newcommand{\MPhf}[1]{\MPworker{purple}{#1}}
    \newcommand{\SP}[1]{\MPworker{olive}{STVS: #1}}
\else
   \newcommand{\MP}[1]{}
   \newcommand{\MPla}[1]{}
    \newcommand{\MPtg}[1]{}
    \newcommand{\MPhf}[1]{}
    \newcommand{\SP}[1]{}
\fi
\newcommand\rmv[1]{}

\newcommand{\techReportOnly}[1]{}

\newcommand{\providerfunctions}{value-added functions\xspace}

\newcommand{\interestextraction}{\interest extraction\xspace}

\newcommand{\topicextraction}{\interestextraction}

\newcommand{\interests}{\interest{s}\xspace}
\newcommand{\interest}{topic\xspace}

\newcommand{\xpirname}{\mbox{\textsc{xpir-bv}}\xspace}
\newcommand{\lingspam}{Ling-spam\xspace}
\newcommand{\binaryNB}{GR-NB\xspace}

\newcommand{\hybrid}{\text{Baseline}\xspace}

\newcommand{\nopriv}{\text{NoPriv}\xspace}

\newcommand{\gen}{{\small\textsf{Gen}}\xspace}
\newcommand{\enc}{{\small\textsf{Enc}}\xspace}
\newcommand{\dec}{{\small\textsf{Dec}}\xspace}

\newcommand{\given}{\,|\,}

\newcommand{\bin}{b_\text{in}}
\newcommand{\fin}{f_\text{in}}

\newcommand{\circledone}{\ding{192}\xspace}
\newcommand{\circledtwo}{\ding{193}\xspace}
\newcommand{\circledthree}{\ding{194}\xspace}
\newcommand{\circledfour}{\ding{195}\xspace}

\newcommand{\cpu}{\textsc{cpu}\xspace}

\def\compactify{\itemsep0in \topsep2pt \parsep=0.00in \partopsep=0pt
\leftmargin2em}
\let\latexusecounter=\usecounter

\newenvironment{myenumerate}
  {\def\usecounter{\compactify\latexusecounter}
   \begin{enumerate}}
  {\end{enumerate}\let\usecounter=\latexusecounter}

\newenvironment{myenumerate2}
  {\def\usecounter{\itemsep=0ex \topsep1ex \parsep=1ex \partopsep=0pt
\leftmargin\parindent\latexusecounter}
   \begin{enumerate}}
  {\end{enumerate}\let\usecounter=\latexusecounter}

\newenvironment{myitemize}%
  {\begin{list}{\labelitemi}{\itemsep3pt \topsep3pt \parsep0.00in
  \partopsep=3pt \leftmargin1em}}%
  {\end{list}}

  {\begin{list}{\labelitemi}{\itemsep6pt \topsep6pt \parsep0.00in
  \partopsep=0pt \leftmargin\parindent}}%
  {\end{list}}

\newenvironment{myitemize3}%
  {\begin{list}{\labelitemi}{\itemsep3pt \topsep3pt \parsep0.00in
  \partopsep=0pt \leftmargin\parindent}}%
  {\end{list}}
  {\begin{list}{\labelitemi}{\itemsep0in \topsep2pt \parsep0.00in
  \partopsep=0pt \leftmargin\parindent}}%
  {\end{list}}

\def\emparagraph#1{\vspace{-1mm}\paragraph{#1}}

\def\discretionaryslash{\discretionary{/}{}{/}}
{\catcode`\/\active
\gdef\URLprepare{\catcode`\/\active\let/\discretionaryslash
        \def~{\char`\~}}}%
\def\URL{\bgroup\URLprepare\realURL}%
\def\realURL#1{\tt #1\egroup}%

\DeclareMathOperator*{\argmax}{argmax}

\begin{document}

\newcommand{\supsyml}[1]{\raisebox{4pt}{\footnotesize #1}}
\newcommand{\rstar}{\supsyml{$\ast$}\xspace}
\newcommand{\rdag}{\supsyml{$\ast\ast$}\xspace}

\author{
          Trinabh Gupta$^{\ast\dag}$\quad
          Henrique Fingler$^{\ast}$\quad
          Lorenzo Alvisi$^{\ast\ddag}$\quad
          Michael Walfish$^{\dag}$ \\[6pt]
\fontsize{9.5}{11}\selectfont $^{\ast}$UT Austin
\quad\quad $^{\dag}$NYU
\quad\quad $^{\ddag}$Cornell
}

\date{}

\title{\textbf{\sys: Email encryption and
provider-supplied functions are compatible}}

\maketitle

\paragraph{Abstract.}
Emails today are often encrypted, but only between mail servers---the
vast majority of emails are exposed in plaintext to the mail servers
that handle them. While better than no encryption, this arrangement
leaves open the possibility of attacks, privacy violations, and other
disclosures. Publicly, email providers have stated that default
end-to-end encryption would conflict with essential %
functions (spam filtering, etc.), because the latter requires analyzing
email text. The goal of this paper is to demonstrate that there
is no conflict. We do so by designing, implementing, and evaluating 
\emph{\sys{}}.
Starting from a cryptographic protocol that
enables two parties to jointly perform a classification task without
revealing their inputs to each other, \sys refines and adapts this
protocol to the email context. Our experimental evaluation of a
prototype demonstrates that email can be encrypted end-to-end \emph{and}
providers can compute over it, at tolerable cost:
clients must devote some storage and processing, and provider overhead
is roughly 5$\times$ versus the status quo.

\section{Introduction}
\label{s:intro}
\label{s:introduction}

Email is ubiquitous and fundamental. For many, it is the principal
communication medium, even with intimates. For these reasons, and
others that we outline below, our animating ideal
in this paper is that email should be \emph{end-to-end private by
  default}.

How far are we from this ideal?  On the plus side, \emph{hop-by-hop}
encryption has brought encouraging progress in protecting email
privacy against a range of network-level attacks. Specifically, 
many emails now
travel between servers over encrypted channels (TLS~\cite{dierks2008transport,
durumeric15neither}).
And network connections between the user and the provider are
often encrypted, for example using HTTPS (in the case of webmail
providers) or VPNs (in the case of enterprise email
accounts).

However, emails are not by default encrypted end-to-end between the
two clients: intermediate hops, such as the sender's
and receiver's provider, handle emails in plaintext. Since
these providers are typically well-run services with a reputation to
protect, many users are willing to just trust them. This trust
however, appears to stem more from shifting \emph{social} norms than
from the fundamental technical safety of the arrangement, which
instead seems to call for greater caution.

Reputable organizations have been known to unwittingly harbor rogue
employees bent on gaining access to user email accounts and other
private user
information~\cite{roguecanadairs,rogueirs,roguegoogle}. If you were
developing your latest startup idea over email, would you be willing
to bet its viability on the assumption that every employee
within the provider acts properly? And well-run organizations are not
immune from hacks~\cite{dnc,sony}---nor, indeed, from the law. Just
in the first half of 2013, Google~\cite{googletransparency},
Microsoft~\cite{microsofttransparency} and Yahoo!~\cite{yahootransparency} collectively
received over 29,000 requests for email data from law enforcement, and
in the overwhelming majority of cases responded with some customer
data~\cite{propublica}.
Many of these requests did not even require a
warrant: under current law~\cite{ecpa}, email older than 180
days can be acquired without judicial approval.  

End-to-end email encryption can shield email contents from prying eyes
and reduce privacy loss when \mbox{webmail} accounts are hacked;
and, while authorities would still be able to
acquire private email by serving subpoenas to the account's owner, 
they would not gain unfettered access to someone's private
correspondence without that party's knowledge. %

Why then are emails not encrypted end-to-end by default?  After all, there
has long been software that implements this functionality,
notably PGP~\cite{zimmermann1995official};
moreover, the large webmail providers offer it as
an option~\cite{google-e2e, yahoo-e2e}
(see also~\cite{somogyi14making, stamos15user, whatsapp, imessagee2e}).
A crucial reason---at least the
one that is often cited~\cite{noe2e1, noe2e2, noe2e3, noe2e4, noe2e5}---%
is that encryption appears to be incompatible with \providerfunctions (such as
spam filtering, email search, and predictive personal
assistance~\cite{googleassistant})
 and with the functions by which ``free'' webmail providers monetize
 user data (for example,
\interestextraction)~\cite{googledatamininghistory}.
These functions are proprietary; for example, the provider might have
invested in training a spam filtering model, and
does not want to publicize it (even if a dedicated party can infer
it~\cite{tramer2016stealing}).
So it follows that the functions must execute on providers' servers
with access to plaintext emails.

But does that truly follow? Our objective in this paper is to refute these claims of
incompatibility, and thus move a step closer to the animating ideal that
we stated at the outset, by
building an alternative, called \sys.

\bigskip

In \sys, senders encrypt email using an end-to-end encryption scheme,
and the intended recipients decrypt and obtain email contents. Then, the
email provider and each recipient engage in a \emph{secure two-party
computation} (2PC); the term refers to cryptographic protocols that
enable one or both parties to learn the output of an agreed-upon
function, without revealing the inputs to each other. For example, a
provider supplies its spam filter, a user supplies an email, and both
parties learn whether the email is spam while protecting the details of
the filter and the content of the email.

The challenge in \sys comes from the 2PC component. There is a tension
between expressive power (the best 2PC schemes can handle any function
and even hide it from one of the two parties) and cost (those schemes
remain exorbitant, despite progress in lowering the costs;
\S\ref{s:yao}). Therefore, in designing \sys, we decided to make certain
compromises to gain even the possibility of plausible performance:
baking in specific algorithms, requiring that the algorithms not be
proprietary (only their inputs, such as model parameters, are hidden),
and incurring per-function design work.

The paper's central example is classification, which \sys applies to
both spam filtering and \interestextraction (for completeness, the
implementation also includes a simple keyword search function).
\sys's first step is to compose
(a)~a relatively efficient 2PC
protocol~(\S\ref{s:hybridbackground}) geared to computations that consist mostly of linear
operations\cite{aditya2016pic, sadeghi2009efficient, evans2011efficient,
blanton11secure, pathak2011privacy}, (b)~linear classifiers from
machine learning (Naive Bayes, SVMs, logistic regression), which fit this form
and which have good accuracy~(\S\ref{s:classification}),
and (c) mechanisms that protect against adversarial
parties.
Although the precise protocol~(\S\ref{s:starting}) has not appeared before,
we don't claim it
as a contribution, as its elements are well-understood.
This combination is simply the jumping-off point for
\sys.

The work of \sys is adapting and incorporating this baseline into a
system for end-to-end encrypted email. In this context, the costs of the
baseline would be, if not quite outlandish, nevertheless too high.
\Sys responds, first, with lower-level protocol refinements: revisiting
the cryptosystem~(\S\ref{s:refinement1}) and designing a packing technique
that conserves calls into the cryptosystem~(\S\ref{s:packing}). Second,
for \topicextraction, \sys rearranges the setup, by decomposing
classification into a non-private step, performed by the
client, which prunes the set of topics; and a
private step that further refines this candidate set to a single topic. Making
this work requires a modified protocol that, roughly speaking, selects a
candidate maximum from a particular subset, while hiding that
subset~(\S\ref{s:pruning}). Third, \sys applies well-known ideas
(feature selection to reduce costs, various mechanisms to guard against
misuses of the protocol, etc.); here, the work is demonstrating that
these are suitable in the present context.  

We freely admit that not all elements of \sys are individually
remarkable.  However, taken together, they produce the first (to our
knowledge) demonstration that classification can be done privately, at
tolerable cost, in the email setting.

Indeed, evaluation~(\S\ref{s:eval}) of our
implementation~(\S\ref{s:impl}) indicates that \sys's cost, versus a
legacy non-private implementation, is estimated to be up to $5.4\times$,
with additional client-side requirements of several hundred megabytes of
storage and per-email \cpu cost of several hundred
milliseconds. These costs represent reductions versus the starting
point~(\S\ref{s:starting}) of up to $100\times$.

Our work here has clear limitations~(\S\ref{s:discussion}). Reflecting
its prototype status, our implementation does not hide metadata, and
handles only the three functions mentioned (ideally, it would handle
predictive personal assistance, virus scanning, and more). More
fundamentally, \sys compromises on functionality; by its design, both
user and provider have to agree on the algorithm, with only the inputs
being private. Most fundamentally, \sys cannot achieve the ideal of
perfect privacy; it seems inherent in the problem setup that one party
gains information that would ideally be hidden. On the
other hand, these leaks are generally bounded, and concerned users can
opt out, possibly at some dollar cost~(\S\ref{s:maliciousparties}, \S\ref{s:discussion}).

The biggest limitation, though, is that \sys cannot change the world on
its own.  As we discuss later~(\S\ref{s:obstacles}), there are other
obstacles en route to the ultimate goal: %
general deployment difficulties, key management, usability,
and even politics. However, we hope that the exercise of working through
the technical details to produce an existence proof (and a rough cost
estimate) will
at least 
shape discourse about the viability of default end-to-end email encryption.
\section{Architecture and overview of \sys}
\label{s:overview}
\label{s:design}

\subsection{Design ethos: (non)requirements}
\label{s:designethos}

\Sys would ideally (a)~enable rich computation over email, (b)~hide
the inputs and implementations of those computations, and
(c)~impose negligible overhead. But these three ideals are in tension.
Below we describe the compromises that form \sys's
design ethos.

\begin{myitemize}

\item \emph{Functionality.} We will not insist that \sys provide exact
replicas of the rich computations that today's providers run over emails; instead, the
goal is to retain approximations of these functions.

\item \emph{Provider privacy.} Related to the prior point, \sys will not
support proprietary algorithms; instead, \sys will protect the
\emph{inputs} to the algorithms. For example, all users of \sys will
know the spam filtering model, but the parameters to the model will be
proprietary.

\item \emph{User privacy.} \Sys will not try to enshroud users' email in
complete secrecy; indeed, it seems unavoidable that
computing over emails would reveal some information about them. However,
\sys will be designed to reveal only the
outputs of the computation, and these outputs
will be short (in bits).

\item \emph{Threat model and maliciousness.} \sys will not build in
protection against actions that subvert the protocol's \emph{semantics}
(for example, a provider who follows the protocol to the letter but who
designs the \interestextraction model to try to recover a precise email);
we will deal with this issue by relying on context, a point we elaborate
on later~(\S\ref{s:maliciousparties}, \S\ref{s:discussion}).
\Sys will, however, build in defenses against adversaries that
deviate from the protocol's
\emph{mechanics}; these defenses will not assume particular
misbehaviors, only that adversaries are subject to
normal cryptographic hardness.

\item \emph{Performance and price.} Whereas the status quo imposes
little overhead on email clients, \sys will require storage and
computation at clients. However, \sys will aim to 
limit the storage cost to several hundred megabytes %
and the \cpu cost 
to a few hundred milliseconds of time per
processed email. For the provider, \sys's aim is to limit overhead to small multiples
of the cost in the status quo.

\item \emph{Deployability and usability.} Certain computations, such as
encryption, will have to run on the client; however, \sys will aim to be
configuration-free. Also, \sys must be backwards compatible with
existing email delivery infrastructure (SMTP, IMAP, etc.).
 
\end{myitemize}

\begin{figure}[t!]
\centerline{\includegraphics[width=3.25in]{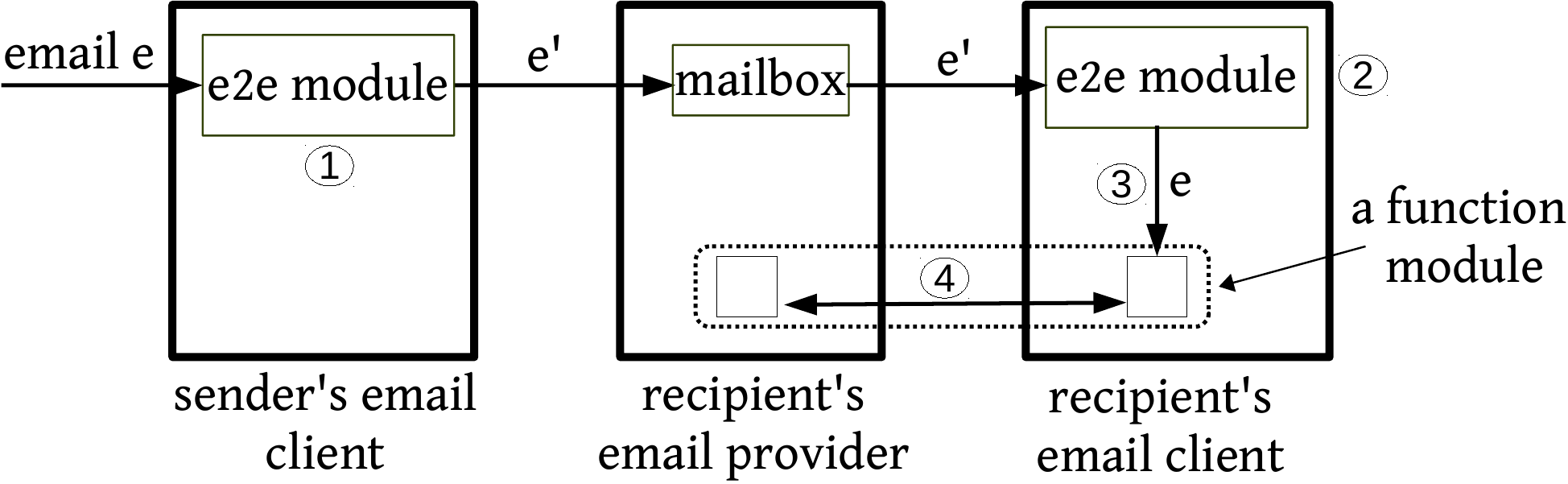}}
\caption{\sys's architecture. $e$ denotes plaintext email; $e'$ denotes
encrypted email.
The sender's provider is not depicted.}
\label{f:arch}
\label{fig:arch}
\end{figure}

\subsection{Architecture}
\label{s:arch}

Figure~\ref{f:arch} shows \sys's architecture. \Sys comprises
an \emph{e2e module} and \emph{function modules}. 
The e2e module implements an end-to-end encryption 
scheme; a function module implements a semantic computation (spam filtering,
etc.).
The e2e module is client-side only, while
a function module has a component at the
client and another at the provider.

At a high level, \sys works as follows.
An email sender uses its e2e module to encrypt and sign an email for
    an email recipient (step~\circledone). The recipient uses its e2e module
    to authenticate and decrypt the email (step~\circledtwo). The e2e module can implement any
    end-to-end encryption scheme; \sys's current prototype uses
    GPG~\cite{gpg}. 
Next, the recipient
    passes the decrypted email contents to the client-side components of the function
modules (step~\circledthree).
Finally, these client-side components 
    participate in a protocol with their counterparts
    at the provider (step~\circledfour).
At the end of the protocol, either the client or the provider learns
the output of the computation.

\sys's e2e module requires cryptographic keys for encrypting,
decrypting, signing, and verifying. Thus, \sys requires a solution to
\emph{key management}~\cite{keybase,whatsapp,melara2015coniks}.
However,
this is a separate effort, deserving of its own paper or product and (as
noted in the introduction) is one of the obstacles that \sys does not
address. Later~(\S\ref{s:discussion}), we discuss why we are optimistic 
that it will ultimately be overcome.

The main work for \sys surrounds the function modules;
the challenge is to balance privacy, functionality,
and performance~(\S\ref{s:designethos}).
Our focus will be on two modules:
  spam filtering and \interestextraction~(\S\ref{s:background}, \S\ref{s:innovations}).
We will also report on a keyword search module~(\S\ref{s:search}).
But before delving into details, we walk through some necessary
    background, on the class of computations run by these modules and 
    the cryptographic protocols that they build on.
\section{Background, baseline, and related work}

\label{s:foundations}
\label{s:background}

\subsection{Classification}
\label{s:classification}

Spam filtering and \interestextraction are classification problems
and, as such, require \emph{classifier algorithms}.
\sys is geared to linear classifiers. %
So far, we have implemented Naive Bayes
(NB)~\cite{metsis2006spam, mccallum1998comparison, graham2002plan,
robinson2003statistical} classifiers, specifically 
a variant of Graham-Robinson's NB~\cite{graham2002plan,
robinson2003statistical}
for spam filtering (we call this variant \binaryNB),\footnote{The original Graham-Robinson NB protects against spam emails
that hide a short message within a large non-spam
text~\cite{grahambetter}. We do not implement that piece; the resulting change
in classification accuracy is small (\S\ref{s:spameval}).}
 and multinomial NB~\cite{mccallum1998comparison} for
\interestextraction;
Logistic Regression (LR) classifiers~\cite{goodman2006online,
ng2002discriminative, lin2008trust, fan2008liblinear},
specifically binary LR~\cite{lin2008trust} and multinomial
LR~\cite{fan2008liblinear}
for spam filtering and
\interestextraction respectively; and 
linear Support Vector Machine (SVM) classifiers~\cite{cortes1995, joachims1998text,
sculley2007relaxed, bosersvm}, 
specifically two-class and one-versus-all SVM~\cite{bosersvm} for spam filtering 
and \interestextraction respectively.
These algorithms, or variants of them, 
yield high 
accuracy~\cite{cormack07trec2007, graham2002plan, huang2003comparing,
zhang2004optimality, joachims1998text, goodman2006online}
(see also \S\ref{s:spameval}, \S\ref{s:topiceval}),
and are used both
commercially~\cite{aberdeen2010learning} and in popular
open-source software
packages for spam filtering, classification, and general machine
learning~\cite{spamprobe,spambayes,spamassassin,scikit-learn,weka,
fan2008liblinear}.
The three types of classifiers differ in their underlying assumptions and 
how they learn parameters from training data. However, when
applying a trained model, they 
all perform analogous linear operations.
We will use Naive Bayes %
as a running example, because it is the simplest
to explain.

\paragraph{Naive Bayes classifiers.}

These algorithms assume that a document can belong to one of several
\emph{categories} (for example, spam or non-spam). The
algorithms output a prediction of a document's category.

Documents (emails) are represented by \emph{feature vectors} $\vec{x}
{=} (x_1, \ldots, x_N)$, where $N$ is the total number of features.  A
feature can be a word, a group of words, or any other efficiently
computable aspect of the document; the algorithms do not assume a
particular mapping between documents and feature vectors, only that some
mapping exists. 
In the \binaryNB spam
classifier~\cite{graham2002plan,
robinson2003statistical}, $x_i$ is Boolean, and indicates the presence
or absence of feature $i$ in the document; in the multinomial NB
text classifier, $x_i$ is the frequency of feature $i$.

The algorithms take as input a feature vector and a \emph{model} that
describes the categories. A model is a set of vectors $\{(\vec{v}_j,\,
p(C_j)) \}$ $(1 \leq j \leq B)$, where $C_j$ is a category (for example,
spam or non-spam), and $B$ is the number of
categories (two for spam; 2208 for \interests, based on Google's  
public list of \interests~\cite{googleinterests}). %
$p(C_j)$ denotes the assumed a priori category
distribution.
The $i$th entry of $\vec{v}_j$ is denoted $p(t_i \given C_j)$ and 
is, roughly speaking, the probability that feature $i$, call it $t_i$,
appears in documents whose
category is $C_j$.\footnote{In more detail,
the \binaryNB spam classifier assumes that the $\{x_i\}$ are 
realizations of independent, separate Bernoulli random variables (RVs), 
with the probabilities of each RV, $p(t_i \given C_j)$,
depending on the hypothesized category. The multinomial NB text classifier
assumes that the $\{x_i\}$ follow a multinomial distribution, with $N$ bins and $\sum_i
x_i$ trials, where the bin probabilities are $p(t_i \given C_j)$ and depend on 
the hypothesized category.}

The \binaryNB spam classification algorithm labels an email, as represented by
feature vector $\vec{x}$, as spam if $p(\textrm{spam}\given \vec{x})$ is
greater than some fixed threshold. To do so, the algorithm computes
\mbox{$\alpha=1/p(\textrm{spam} \given \vec{x}) - 1$} in log space. One can
\ifx\buildtechreport\undefined
show~\cite[Apdx~A.1]{self}
\else
show~(Apdx~A.1)
\fi
that $\log \alpha$ is
equivalent to:
\begin{align}
\nonumber &\left(\sum_{i=1}^{i=N} x_i \cdot \log{p(t_i \given C_2)}\right) + 1
\cdot \log {p(C_2)} \\ 
- &\left(\sum_{i=1}^{i=N} x_i \cdot \log {p(t_i \given C_1)}\right) + 1
\cdot \log {p(C_1)},
\label{eq:dot1}
\end{align}
where $C_1$ represents spam and $C_2$ represents non-spam.

For the multinomial NB text classifier, selection works by identifying the
category $C_{j^*}$ that maximizes likelihood: $j^* = \text{argmax}_j\,
p(C_j \given \vec{x})$.
One can 
\ifx\buildtechreport\undefined
show~\cite[Apdx A.2]{self} 
\else
show~(Apdx A.2)
\fi
that it suffices
to select the $C_j$ for which the following is maximal:
\begin{align}
&\left(\sum_{i=1}^{i=N} x_i \cdot \log{p(t_i \given C_j)}\right) + 1
\cdot \log {p(C_j)}.
\label{eq:dot2}
\end{align}

For LR and SVM classifiers, the term $\log{p(t_i \given C_j)}$ is
replaced by a ``weight'' term $w_{i,j}$ for feature $x_i$ and category
$C_j$, and $\log{p(C_j)}$ is replaced by a ``bias'' term $b_j$ for category $j$.

\subsection{Secure two-party computation}

To perform the computation described above within a
function module (\S\ref{s:arch}) securely, i.e., in a way that the client does not learn
the model parameters and the provider does not learn the feature vector,
\sys uses \emph{secure two-party computation} (2PC):
cryptographic protocols that enable two parties to
compute a function without revealing their inputs to each
other~\cite{yao1982protocols, goldreich1987play}.
\Sys
builds on a relatively efficient 2PC protocol~\cite{aditya2016pic,
evans2011efficient, sadeghi2009efficient, blanton11secure,
pathak2011privacy} that we name
\textbf{\yaogllmname};
we present this below, informally and bottom up.
(For details and rigorous
descriptions, see~\cite{lindell2009proof,
huang11faster, goethals2004private, sadeghi2009efficientfullpaper}.)

\begin{figure*}[t]
\hrule
\medskip
{\fontsize{9}{10.8}\selectfont

\begin{center}
\textbf{\yaogllmname}
\end{center}

\begin{myitemize3}
\item The protocol has two parties. Party A begins with a
matrix; Party B begins with a vector. The protocol computes a
vector-matrix product and then performs an arbitrary computation, $\phi$, on
the resulting vector; neither party's input is revealed to the other.

\item The protocol assumes an additively homomorphic encryption
(AHE) scheme
    $(\gen,\enc,\dec)$, meaning that
    $\enc(pk, m_1) \cdot \enc(pk, m_2) {=} \enc(pk, m_1 + m_2)$, 
     where $m_1, m_2$ are plaintext messages, %
    $+$ represents addition of two plaintext messages, and $\cdot$ is an
    operation on the ciphertexts. This also implies
    that given a constant $z$ and $\enc(pk, m_1)$, one can compute
    $\enc(pk, z \cdot m_1)$. %
\end{myitemize3}

\textbf{Setup phase}
\begin{myenumerate}
\item Party A forms a matrix with columns $\vec{v}_1,\ldots,\vec{v}_B$;
each vector has $N$ components. It does the following:
\begin{myitemize3}
        \item Generates public and secret keys
        $(pk,sk)\gets\gen(1^{k})$, where $k$ is a security parameter.

        \item \label{e:encryptmodel} Encrypts each column component-wise, so
        $\enc(pk, \vec{v}_j) = (\enc(pk, v_{1,j}), \ldots, \enc(pk, v_{N,j}))$.

        \item Sends the encrypted matrix columns and $pk$ to Party B.
\end{myitemize3}

\end{myenumerate}

\vspace{0.5ex}

\textbf{Computation phase}
\begin{myenumerate}
\setcounter{enumi}{1}
\item Party B begins with an $N$-component
vector $\vec{x}{=}(x_1, \ldots, x_N)$. It does the following:

    \begin{myitemize3}
        \item \label{e:dotproducts} (dot products) Computes encrypted dot product
        for each
        matrix column: 
        $\enc(pk, d_j) {=} \enc(pk, \sum_{i=1}^N x_i \cdot v_{i,j})$,
        this abuses notation, since the encryption function is not
        deterministic. The computation relies on the homomorphic
        property.

        \item \label{e:blinding} (blinding) Blinds $d_j$ by adding random noise $n_j \in_R \{0,
          1\}^{b+\delta}$. 
        That is, computes $\enc(pk, d_j + n_j) {=} \enc(pk, d_j) \cdot
        \enc(pk, n_j)$. Here $b$ is the bit-length of $d_j$ and $\delta \geq 1$ is
        a security parameter. %

        \item \label{e:encnoise} Sends $(\enc(pk, d_1 + n_1), \ldots,
        \enc(pk, d_B + n_B))$ to Party A.
    \end{myitemize3}

\item \label{e:providerdecrypt} Party A applies $\dec$ component-wise, to get $(d_1 + n_1, \ldots, d_B + n_B)$

\item \label{e:yao} %
    Party A and Party B participate in Yao's 2PC protocol; they use a
    function $f$ that subtracts the
    noise $n_j$ from $d_j + n_j$ and applies the function $\phi$ to the $d_j$.
    One of the two parties (which one depends on the arrangement)
    obtains the output $\phi(d_1,\ldots,d_B)$.
\end{myenumerate}

}
\hrule
\caption{\yaogllmname.
This protocol~\cite{aditya2016pic, evans2011efficient, sadeghi2009efficient,
blanton11secure, pathak2011privacy} combines GLLM's secure dot products~\cite{goethals2004private}
with Yao's general-purpose 2PC~\cite{yao1982protocols}.
\Sys's design and implementation apply this protocol to the linear
classifiers described in \S\ref{s:classification}. 
The provider is Party A, and the client is Party
B.
\Sys's instantiation of this protocol incorporates several additional
elements~(\S\ref{s:starting}): a variant of 
Yao~\cite{keller2015actively, ishai2003extending} that defends 
against actively adversarial parties; amortization of the expense of this
variant via precomputation in the setup phase;
a technique to defend against adversarial key generation (for
example, not invoking \gen); and a packing technique~(\S\ref{s:packing})
in
step~\ref{e:encryptmodel} (bullet 2) and step~\ref{e:dotproducts}
(bullet 1). 
}
\label{f:gllm+yao}
\end{figure*}

\paragraph{Yao's 2PC.}
\label{s:yao}
\label{s:backgroundmpc}
A building block of \yaogllmname is the classic scheme of
Yao~\cite{yao1982protocols}. 
Let $f$ be a function, represented as a Boolean circuit (meaning a
network of Boolean gates: AND, OR, etc.), with $n$-bit input, and let there be
two parties $P_1$ and $P_2$ that supply separate pieces of this input, denoted
$x_1$ and $x_2$, respectively. 
Then Yao (as the protocol is sometimes known), when run between $P_1$
and $P_2$, takes as inputs $f$ and
$x_1$ from $P_1$, $x_2$ from $P_2$, and outputs $f(x_1, x_2)$ to $P_2$, such
that neither party learns about the other's input: $P_1$ does not learn anything
about $x_2$, and $P_2$ does not learn anything about $x_1$ 
except what can be inferred from %
$f(x_1, x_2)$. 

At a very high level, Yao works by having one party generate
encrypted truth tables, called \emph{garbled Boolean gates}, for gates in the
original circuit, and having the other party decrypt and thereby evaluate
the garbled gates.

In principle, Yao handles arbitrary functions. In practice,
however, the costs are high. A big problem is the computational model.
For example, 32-bit multiplication, when represented as a Boolean
circuit, requires on the order of 2,000 gates, and each of those gates
induces cryptographic operations (encryption, etc.). %
Recent activity has improved the costs
(see~\cite{%
zahur2015two, %
huang2012quid, %
keller2015actively, %
lindell2016fast, %
kreuter2012billion, %
songhori2015tinygarble, %
gordon2012secure, zahur2016revisiting} %
and references therein), but the bottom line is still too expensive to
handle arbitrary computations.
Indeed, \sys's prototype uses Yao very selectively---just
to compute several comparisons of 32-bit numbers---and even then it turns
out to be a
bottleneck~(\S\ref{s:spameval}, \S\ref{s:topiceval}), despite using
a recent and optimized %
implementation~\cite{zahur2015obliv, zahur2015two}.

\paragraph{Secure dot products.}
\label{s:sspbackground}

Another building block of \yaogllmname is a secure dot
product protocol, specifically GLLM~\cite{goethals2004private}.
Many such protocols (also called secure scalar product (SSP)
protocols) have been proposed~\cite{atallah2001secure,
du2001privacy,
ioannidis2002secure,
du2001privacy2,
du2001protocols,
vaidya2002privacy,
amirbekyan2007new,
trincua2007fast,
du2002practical,
xu2010research,
zhu2015fast,
zhu2012efficient,
zhu2015efficient,
du2002building,
shaneck2010efficient,
goethals2004private,
dong2014fast,
wright2004privacy}. %
They fall into two categories:
    those that are provably secure~\cite{goethals2004private,
dong2014fast, wright2004privacy} and 
    those that either have no security proof or
    require trusting a third party~\cite{atallah2001secure,
    du2001privacy,
    ioannidis2002secure,
    du2001privacy2,
    du2001protocols,
    vaidya2002privacy,
    amirbekyan2007new,
    trincua2007fast,
    du2002practical,
    xu2010research,
    zhu2015fast,
    zhu2012efficient,
    zhu2015efficient,
    du2002building,
    shaneck2010efficient}. 
Several protocols in the latter category have been attacked~\cite{laur2004private,
kang2011fast,
huang2005privacy,
chiang2005secrecy,
goethals2004private}.
GLLM~\cite{goethals2004private}
is in the first category, is
state of the art, and is widely used. 

\paragraph{Hybrid: \yaogllmname.}
\label{s:hybridbackground}

\Sys's starting point is \yaogllmname, a hybrid of Yao and GLLM.  
It is depicted in Figure~\ref{f:gllm+yao}.
One party starts with a matrix, and encrypts the entries. 
The other party starts with a vector and leverages \emph{additive} (not
fully)
homomorphic encryption (AHE) to (a)~compute the vector-matrix product in cipherspace,
and (b)~blind the resulting vector. The first party then decrypts to obtain the
blinded vector. The vector then feeds into Yao: the two
parties remove the blinding and perform some computation $\phi$.

\yaogllmname has been applied to spam filtering using
LR~\cite{pathak2011privacy}, face recognition using
SVM~\cite{aditya2016pic}, and 
face and biometric identification using Euclidean
distance~\cite{evans2011efficient,
blanton11secure, sadeghi2009efficient}.

\paragraph{Other related work.} 
There are many works on private classification that do not build
on \yaogllmname.
They rely on
alternate
building blocks or hybrids:
additively homomorphic encryption~\cite{bost2014machine,
liu2016privacy} (AHE), fully homomorphic
encryption~\cite{khedr2014shield} (FHE),
or a different Yao hybrid~\cite{bringer2016boosting}.
For us, \yaogllmname appeared to be a more promising
starting point (as examples, \yaogllmname
contains a packing optimization that
significantly saves client-side storage resources relative
to~\cite{bost2014machine}, and in contrast to~\cite{khedr2014shield},
\yaogllmname reveals only the final output of the
computation rather than intermediate dot products). %

Another related line of research focuses on privacy and linear
classifiers---but in the \emph{training} phase.
Multiple parties can train a global model without revealing their
private inputs~\cite{kantarcioglu2003privacy,
yi2009privacy,
sumana2014privacy,
gangrade2012privacy,
zhan2008privacy,
yang2005privacy,
vaidya2004privacy,
gangrade2013privacy,
zhan2005privacy,
vaidya2008privacy,
zhan2004privacy,
vaidya2008privacySVM,
yu2006privacyhorizontal,
yu2006privacyvertical}, or a party can release a trained ``noisy'' model
that hides its training data~\cite{zhang2005privacy,
kaleli2007providing,
vaidya2013differentially}.
These works are complementary to \sys's focus on \emph{applying} the
model. %

\subsection{Baseline protocol}
\label{s:starting}

\Sys begins by applying the \yaogllmname
protocol~(Figure~\ref{f:gllm+yao}, \S\ref{s:hybridbackground})
to the algorithms described in Section~\ref{s:classification}.
This works because expressions~\eqref{eq:dot1}
and~\eqref{eq:dot2} are dot products of the necessary form.
Specifically,
the provider is party A and supplies
$(\vec{v}_j,\,p(C_j))$, the client is party B and
supplies $(\vec{x},\,1)$, and the protocol computes their dot product.
Then, the threshold comparison (for spam filtering) or the maximal selection (for
\interestextraction) happens inside an instance of Yao.
For spam filtering, the client receives the classification output;
    for \interestextraction, the provider does.
Note that storing the encrypted model at the client 
is justified 
by an assumption that model vectors change infrequently.

In defining this baseline, we include mechanisms to defend against
adversarial parties~(\S\ref{s:arch}).  Specifically, whereas under the
classical Yao protocol an actively adversarial party can obtain the
other's private inputs~\cite{ishai2003extending}, \sys incorporates a
variant~\cite{ishai2003extending, keller2015actively} that solves this
problem.  This variant brings some additional expense, but it can be
incurred during the setup phase and amortized. Also, \yaogllmname
assumes that the AHE's key generation is done honestly, whereas we would
prefer not to make that assumption; \sys incorporates the standard
response.\footnote{In more detail, the AHE has public parameters which,
if chosen adversely (non-randomly) would undermine the expected
usage. To get around this, \sys determines these parameters with
Diffie-Hellman key exchange~\cite{merkle78secure, diffie1976new}, 
so that both parties inject randomness into these parameters.}

While the overall baseline is literally new (\yaogllmname was previously
used in weaker threat models, etc.), its elements are well-known, so we
do not claim novelty.

\newcommand{\pail}{pail}
\newcommand{\xpir}{xpir}
\newcommand{\epail}{e_\text{\pail}}
\newcommand{\cpail}{c_\text{\pail}}
\newcommand{\apail}{a_\text{\pail}}
\newcommand{\dpail}{d_\text{\pail}}
\newcommand{\ppail}{p_\text{\pail}}
\newcommand{\expir}{e_\text{\xpir}}
\newcommand{\cxpir}{c_\text{\xpir}}
\newcommand{\axpir}{a_\text{\xpir}}
\newcommand{\dxpir}{d_\text{\xpir}}
\newcommand{\pxpir}{p_\text{\xpir}}
\newcommand{\otbase}{\textit{bot}_\text{base}}
\newcommand{\otbasegen}{\textit{bot}_\text{gen}}
\newcommand{\otbaseeval}{\textit{bot}_\text{eval}}
\newcommand{\otbaseprov}{\textit{bot}_\text{prov}}
\newcommand{\otbasecli}{\textit{bot}_\text{cli}}
\newcommand{\otext}{\textit{ot}_\text{ext}}
\newcommand{\cpugate}{g}
\newcommand{\gateprov}{\cpugate_\text{prov}}
\newcommand{\gatecli}{\cpugate_\text{cli}}
\newcommand{\gategen}{\cpugate_\text{gen}}
\newcommand{\gateeval}{\cpugate_\text{eval}}
\newcommand{\cpuyao}{y_\text{per-in}}

\newcommand{\szgate}{\textit{sz}_\text{gate}}
\newcommand{\szval}{\textit{sz}_\text{per-in}}
\newcommand{\szbase}{\textit{sz}_\text{base}}
\newcommand{\szext}{\textit{sz}_\text{ext}}
\newcommand{\szemail}{\textit{sz}_\text{email}}

\newcommand{\btapail}{\beta_\text{\pail}}
\newcommand{\btaxpir}{\beta_\text{\xpir}}
\newcommand{\btaprxpir}{\beta'_\text{xpir}}

\begin{figure*}[t]
\footnotesize
\centering

\medmuskip=1.5mu
\thickmuskip=2mu plus 2mu

\begin{tabular}{
@{}
*{1}{>{\raggedright\arraybackslash}b{.17\textwidth}}  @{ } %
*{1}{>{\raggedright\arraybackslash}b{.15\textwidth}}  @{ } %
*{1}{>{\raggedright\arraybackslash}b{.29\textwidth}}  @{ } %
*{1}{>{\raggedright\arraybackslash}b{.33\textwidth}}  @{ } %
@{}
}

    & Non-private
    & Baseline (\S\ref{s:starting})
    & \Sys (\S\ref{s:refinement1}--\S\ref{s:pruning}) \\

\midrule
\textbf{Setup} \\
Provider \cpu time
    & N/A 
    & $N\cdot\btapail\cdot\epail + K_\text{cpu}$ %
    & $N'\cdot\btaprxpir\cdot\expir + K_\text{cpu}$ %
    \\
Client \cpu time
    & N/A 
    & $K_\text{cpu}$ %
    & $K_\text{cpu}$ %
    \\
Network transfers
    & N/A 
    & $N\cdot\btapail\cdot\cpail + K_\text{net}$%
    & $N'\cdot\btaprxpir\cdot\cxpir + K_\text{net}$%
    \\
Client storage 
    & N/A 
    & $N\cdot\btapail\cdot\cpail$
    & $N'\cdot\btaprxpir\cdot\cxpir$
    \\

\midrule
\textbf{Per-email} \\
Provider \cpu 
    & $L\cdot h + L\cdot B\cdot s$
    & $\btapail\cdot\dpail + B \cdot \cpuyao$ %
    & $\beta''_{\text{xpir}}\cdot\dxpir + B''\cdot \cpuyao$
    \\
Client \cpu 
    & N/A 
    & $L\cdot\btapail\cdot\apail + \btapail\cdot\epail + B \cdot \cpuyao$ %
    & $L\cdot\btaxpir\cdot\axpir + (L+B') \cdot s + \beta''_{\text{xpir}}\cdot\expir + B''\cdot \cpuyao$
    \\
Network 
    & $\szemail$
    & $\szemail$ + $\btapail\cdot\cpail + B \cdot \szval$%
    & $\szemail$ + $\beta''_{\text{xpir}}\cdot\cxpir + B''\cdot \szval$
    \\
\bottomrule
\end{tabular}

\begin{tabular}{l l}
$L$ = number of features (\S\ref{s:starting}) in an email &
$h$  = \cpu time to extract a feature and lookup its conditional
probabilities %
\\
$B$ = number of categories in the model (\S\ref{s:starting}) &

$s$  = \cpu time to add two probabilities \\ %
$\szemail$ = size of an email & 
$N'$ = number of features selected after aggressive feature selection
(\S\ref{s:pruning}) \\
$N$ = number of features in the model (\S\ref{s:starting}) &
$e$ = encryption (in an AHE scheme) \cpu time  \\
$d$ = decryption (in an AHE scheme) \cpu time   &
$c$ = ciphertext size in an AHE scheme \\
$a$ = homomorphic addition (in an AHE scheme) \cpu time (Fig.~\ref{f:gllm+yao}) &
$b$ = $\log L + \bin + \fin$ (\S\ref{s:packing}) \\
$\bin$ = \# of bits to encode a model parameter (\S\ref{s:packing})  &
$\fin$ = \# of bits for the frequency of a feature in an email
(\S\ref{s:packing}) \\
$p$ = \# of $b$-bit probabilities packed in a ciphertext (\S\ref{s:packing}) &
$s$ = ``left-shift'' \cpu time in \xpirname (\S\ref{s:packing}) \\
$B''$ = $B$ (if spam) or $B'{=}$ \# of candidate topics ($\ll B$) (\S\ref{s:pruning}) (if topics) &
$\beta_\text{pail}:=\ceil{B/p_\text{pail}},\beta_\text{xpir}:=\ceil{B/p_\text{xpir}}$  \\
$\beta'_\text{xpir}:=\floor{B/p_\text{xpir}} + 1/\floor{p_\text{xpir}/k}$ 
where $k{=}(B \bmod p_\text{xpir})$ &
$\beta''_\text{xpir}:= \beta_{\text{xpir}}$ (if spam) or $B'$ (if topics) \\
$\cpuyao, \szval$ = Yao \cpu time and network transfers
per $b$-bit input (\S\ref{s:backgroundmpc}) &
$K_\text{cpu}, K_\text{net}$ = constants for \cpu and network costs
(\S\ref{s:starting}) \\

\end{tabular}

\normalfont\selectfont
\caption{%
Cost estimates for classification. Non-private refers to a system in
which a provider locally classifies plaintext email. The baseline is
described in Section~\ref{s:starting}. 
Microbenchmarks are given in \S\ref{s:microeval}.}
\label{fig:costs}
\label{f:costs}
\end{figure*}

\section{\Sys's protocol refinements}
\label{s:innovations}

The baseline just described is a promising foundation. But adapting it
to an end-to-end system for encrypted email requires work. The main
issue is costs.  As examples, for a spam classification model with
$N{=}5$M features, the protocol consumes over 1~GB of client-side
storage space; for \interestextraction with $B{=}2048$ categories, it
consumes over 150~ms of provider-side \cpu time and 8~MB in network
transfers (\S\ref{s:eval}). Another thing to consider is the robustness of the
guarantees.

This section describes \sys's refinements, adjustments, and
modifications. The nature of the work varies from low-level
cryptographic optimizations, to architectural rearrangement, to
applications of known ideas (in which case the work is demonstrating
that they are suitable here). We begin with refinements that are aimed
at reducing costs~(\S\ref{s:refinement1}--\S\ref{s:pruning}), the
effects of which are summarized in Figure~\ref{f:costs}; then we
describe \sys's robustness to misbehaving
parties~(\S\ref{s:maliciousparties}).

\subsection{Replacing the cryptosystem}
\label{s:refinement1}

Both \sys's protocol and the baseline require additively homomorphic
encryption~(Figure~\ref{f:gllm+yao}). The traditional choice for
AHE---%
it is used in prior works~\cite{aditya2016pic, evans2011efficient, sadeghi2009efficient,
pathak2011privacy}---is Paillier~\cite{paillier99public}, which is based on a
longstanding number-theoretic presumed hardness assumption.
However, Paillier's \dec takes hundreds of microseconds
on a
modern CPU, which contributes substantially
to provider-side \cpu time. %

Instead, \sys 
turns to a
cryptosystem based on the \emph{Ring-LWE
assumption}~\cite{lyubashevsky2010ideal}, %
a relatively young assumption (which is 
usually a disadvantage in cryptography) but one that
has nonetheless received a lot of recent
attention~\cite{brakerski2011fully,lyubashevsky2013toolkit,surveyrlwemsr,liu2015efficient}.
Specifically, \sys 
incorporates the additively homomorphic cryptosystem of
Brakerski and Vaikuntanathan~\cite{brakerski11fully}, %
as implemented and optimized by Melchor et
al.~\cite{melchor2016xpir} in the XPIR system;
we call this \xpirname. %
This change brings the cost of each invocation of $\dec$
down over an order of magnitude, to scores of
microseconds~(\S\ref{s:microeval}), and similarly with
$\enc$.
The gain is reflected in the cost model (Figure~\ref{f:costs}),
in replacing $\dpail$ with $\dxpir$ (likewise with
$\epail$ and $\expir$, etc.)

However, the change makes ciphertexts 64$\times$ larger: from 256~bytes
to 16~KB. Yet, this is not the disaster that it seems. Network costs do
increase
(in Figure~\ref{f:gllm+yao}, step~\ref{e:encnoise}, bullet 3),
but by far less than 64$\times$. Because the \emph{domain} of the encryption
function grows, one can tame what would otherwise be
an explosion in network and storage, and also gain further \cpu
savings. We describe this next. %

\subsection{Packing}
\label{s:packing}

The basic idea is to represent
multiple plaintext elements in a single ciphertext; this opportunity
exists because the domain of $\enc$ is much larger than any single
element that needs to be encrypted. Using packing, one can reduce the
number of invocations of $\enc$ and $\dec$ in Figure~\ref{f:gllm+yao},
specifically in
step~\ref{e:encryptmodel} bullet 2,
step~\ref{e:blinding} bullet 2, and 
step~\ref{e:providerdecrypt}.
The consequence is a significant reduction in resource consumption,
specifically provider \cpu time 
and storage
for spam filtering, and provider \cpu time
for \interestextraction.

The challenge 
is to preserve the semantic
content of the underlying operation (dot product) within cipherspace.
Below, we describe prior work, and then how \sys overcomes
a limitation in this work.

\emph{Prior work.}
A common packing technique is as follows.
Let $G$ denote number of bits in the domain of the encryption algorithm $\enc$,
$\bin$ denote the number of semantically useful
  bits in an element that would be encrypted (a model parameter in our
  case), and $\fin$ denote the number of bits for the multiplier of an
encrypted element (frequency of a feature extracted from an email in our case).
The output of a dot product computation---assuming a sum of $L$
products, each
formed from a $\bin$-bit element and a $\fin$-bit element---has $b=\log L +
  \bin + \fin$
``semantic bits,''
(in our context, $L$ would be the number of features extracted from an
email).
This means that there is ``room'' to pack $p=\ceil{G/b}$ bits
into a single ciphertext. 

GLLM~\cite{goethals2004private} incorporates this technique; this takes
place in Figure~\ref{f:gllm+yao}, step~\ref{e:encryptmodel}, bullet
2 (though it isn't depicted in the figure).
In that step, GLLM
traverses each row in its matrix from
left to right, encrypting together sets of $p$ numbers. If fewer than
$p$ numbers are left, the scheme encrypts those together but
does not cross a row.
Then, in step~\ref{e:dotproducts}, bullet 1 it performs dot products on the packed
ciphertexts, by exploiting the fact that the elements that need to be
added are aligned.
For example, if the elements in the first row ($v_{1, 1}, \ldots,
v_{1,p}$) are to
be added to those in the second row ($v_{2, 1}, \ldots, v_{2,p}$), then 
the ciphertext space operation  applied to
$c_1 {=} \enc(pk, v_{1,1} \| \ldots \| v_{1,p})$ 
and $c_2 {=} \enc(pk, v_{2,1} \| \ldots \| v_{2,p})$ yields
$c_3 = c_1 \cdot c_2 = \enc(pk, v_{1,1} + v_{2,1} \| \ldots \| v_{1,p} +
v_{2,p})$. For this to work, the individual sums (e.g., $v_{1,p}
+ v_{2,p}$) cannot overflow $b$ bits.

\begin{figure}[t!]
\centerline{\includegraphics[width=3.25in]{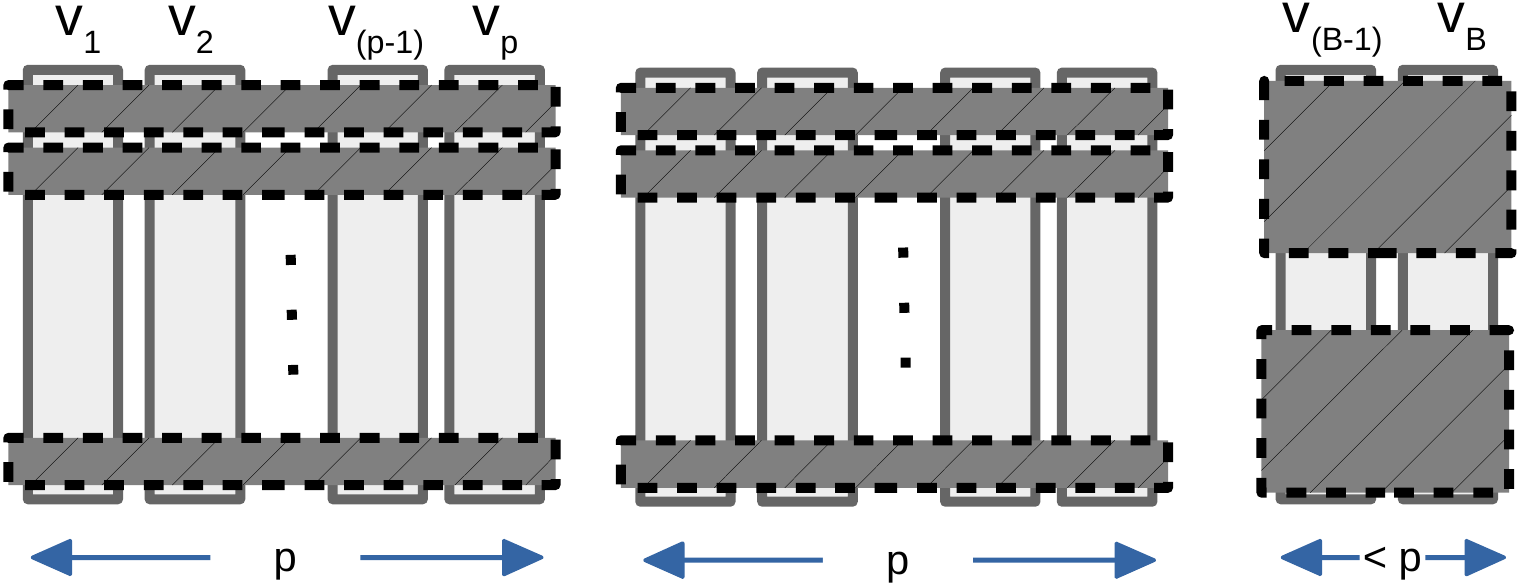}}
\caption{Packing in \sys.}
\label{f:packing}
\label{fig:packing}
\end{figure}

\emph{\Sys's packing scheme.} With the technique above, the ``last
ciphertext'' in a row contains extra space whenever $B$ is larger than a
multiple of $p$; this is because ciphertexts from different rows are not
packed together.  If $\ceil{B/p}$ is large, then this waste is
insignificant.  However, if $B$ is small---and in particular is much
smaller than $p$---then the preceding technique leaves a substantial
optimization opportunity on the table.  For example, if $B=2$ (as in the
spam filtering application) and $p=1024$ (as in the \xpirname
cryptosystem)
then the technique described above packs two elements per ciphertext, gaining a
factor of two, when a factor of 1024 was possible in principle.

\begin{figure*}[t]
\hrule
\medskip
{\fontsize{9}{10.8}\selectfont

\newcommand{\listpackedcts}{\vec{\textit{pcts}}}
\newcommand{\mapping}{\textit{cat2col}}
\newcommand{\mappingtwo}{\textit{feat2row}}
\newcommand{\listextracted}{S'}

\begin{center}
\textbf{\sys's protocol for proprietary \interestextraction, based on
candidate topics}
\end{center}

\begin{myitemize3}
\item 
The protocol has two parties. Party A begins with a matrix $\vec{v}_1,
\ldots, \vec{v}_{B}$.
Party B begins with a vector 
$\vec{x}{=}(x_1, \ldots, x_{N})$ and a
list $\listextracted$ of $B' < B$ column indexes, where each index is
between $1$ and $B$; $\listextracted$ indicates a subset of the columns
of matrix $\vec{v}$.
The protocol constructs a vector from the 
product of $\vec{x}$ and the 
submatrix of $\vec{v}$ given by $\listextracted$,
and outputs the column index (in $\vec{v}$) that 
corresponds to the maximum element in the vector-submatrix product;
neither party's input is revealed to the other.

\item The protocol has two phases: setup and computation. The setup phase 
is as described in
Figure~\ref{f:gllm+yao} but with 
the addition of packing
from \S\ref{s:packing}. 

\end{myitemize3}

\textbf{Computation phase}
\begin{myenumerate}
\setcounter{enumi}{2}
\item 
Party B 
does the following:

    \begin{myitemize3}

        \item (compute dot products) \label{e:prundotproducts}
        As described
        in Figure~\ref{f:gllm+yao}, step~\ref{e:dotproducts}, bullet 1, and
        \S\ref{s:packing}. %
        At the end of the 
        dot product computations,
        it gets a vector of packed ciphertexts
        $\listpackedcts{=}(\enc(pk, d_1 \| \ldots \| d_p), \ldots, \enc(pk, \ldots \|
        d_B \| \ldots))$,
        where
        $d_i$ is the dot product of $\vec{x}$ and the $i$-th matrix column $\vec{v}_i$, and $p$ is the
        number of $b$-bit positions in a packed ciphertext (\S\ref{s:packing}).
        
        \item \label{e:leftshifts} (separate out dot products for the columns in
        $\listextracted$ from the rest) 
        For each entry in $\listextracted$, i.e.,
        $\listextracted[j]$,
        makes a copy of the packed ciphertext containing 
        $d_{\listextracted[j]}$, and shifts 
        $d_{\listextracted[j]}$
        to the left-most $b$-bit position in that ciphertext.
        Because each ciphertext holds $p$ elements, the
        separation works by using the quotient and remainder of
        $\listextracted[j]$, when divided by $p$, to identify, respectively, the relevant
        packed ciphertext and position within it.
        That is, for $1\leq j \leq B'$,
        computes ciphertext $%
                \enc(pk, d_{\listextracted[j]} \| \ldots) {=}
                \listpackedcts[Q_j] \cdot 2^{b \cdot R_j}$, where
                $Q_j {=} \ceil{\listextracted[j] / p} - 1$, and 
                $R_j  {=} (\listextracted[j] - 1) \bmod p$.
        The shifting relies on the multiply-by-constant homomorphic
        operation (see Figure~\ref{f:gllm+yao} and \S\ref{s:packing}).

        \item \label{e:blindingandsend} (blinding) Blinds 
        $d_{\listextracted[j]}$
        using the technique described in Figure~\ref{f:gllm+yao},
        step~\ref{e:blinding}, bullet 2, but extended to packed ciphertexts.
        Sends the $B'$ ciphertexts 
                $(\enc(pk, d_{\listextracted[1]} + n_1 \| \ldots),
\ldots, \enc(pk, d_{\listextracted[B']} + n_{B'} \| \ldots))$
        to Party A. Here, $n_j$ is the added noise.

    \end{myitemize3}

\item \label{e:providerdecryptprunedcts} Party A applies \dec on the
$B'$ ciphertexts, followed by bitwise right shift 
on the resulting plaintexts, to get $d_{\listextracted[1]} + n_1, \ldots,
d_{\listextracted[B']} + n_{B'}$. 

\item \label{e:yaowithprunedcts} The two parties engage in Yao's 2PC.
Party B's private inputs are $\listextracted$ and $\{n_j\}$ for $1
\leq j \leq B'$; Party A's
private inputs are 
$\{(d_{\listextracted[j]} + n_j)\}$ for $1 \leq j \leq B'$;
and, the parties use a function $f$ that subtracts $n_j$ from 
$d_{\listextracted[j]} + n_j$, and computes and returns
$\listextracted[\argmax_j d_{\listextracted[j]}]$ %
to Party A. 
\end{myenumerate}

}
\hrule
\caption{%
Protocol for proprietary
    \interestextraction, based on candidate topics (this instantiates
step~(ii) in Section~\ref{s:pruning}).
The provider is Party A;
the client is Party B.
This protocol builds on the protocol presented in
\S\ref{s:starting}--\S\ref{s:packing}.}
\label{f:pruningprotocol}
\label{fig:pruningprotocol}
\end{figure*}

\sys's packing technique ``crosses rows.'' Specifically, \sys splits 
the matrix $\{(\vec{v}_j,\, p(C_j))\}$ into zero or more sets of $p$
column vectors plus up to one set with fewer than $p$
vectors as depicted in Figure~\ref{f:packing}. For the sets with $p$
vectors, \sys follows the technique above.
For the last set, \sys packs elements in row-major order
\emph{without restricting the packing to be within a row}
(see the rightmost matrix in Figure~\ref{f:packing}),
under one constraint: elements in the same row of the matrix must not be put
into different ciphertexts.

The challenge with across-row packing is being able to add elements from
two different rows that are packed in the same ciphertext (in
step~\ref{e:dotproducts}, bullet 1).
To address
this challenge, \sys exploits the homomorphism to cyclically rotate the packed elements in a
ciphertext. %
For example, assume 
$c {=} \enc(pk,\, v_{1, 1} \| \ldots \| v_{1, k} \| 
v_{2, 1} \| \ldots \| v_{2, k})$ 
is a packed ciphertext,
where 
$v_{1, 1}, \ldots, v_{1, k}$ 
are elements from the
first row, and 
$v_{2, 1}, \ldots, v_{2, k}$ 
are from the second row.
To add each 
$v_{1, i}$
with 
$v_{2, i}$ 
for $i \in \{1, \ldots, k\}$, one can ``left-shift''
elements in $c$ by $k$ positions to get 
    $c'{=}\enc(pk, v_{2, 1} \| \ldots \| v_{2, k} \| \ldots)$;
this is done by applying the ``constant multiplication'' operation
(Figure~\ref{f:gllm+yao}, bullet 2), with $z=2^{k\cdot b}$.
At this point, the rows are ``lined up'', and one can operate on $c$ and $c'$ to add the
plaintext elements.

\emparagraph{Cost savings.} Here we give rough estimates of the effect
of the refinements in this subsection and the previous; a more detailed
evaluation is in Section~\ref{s:eval}.
For the spam filtering module,
the provider's \cpu drops by $5\times$ and the client-side storage drops
by $7\times$, relative to the baseline~(\S\ref{s:starting}).
However, \cpu at the client
increases by $10\times$ (owing to the cyclic shifts), 
and the network overhead
increases by $5.4\times$; despite these increases, both
    costs are not exorbitant in absolute terms, and we view them as tolerable
(\S\ref{s:spameval}, \S\ref{s:evaldiscussion}).
The provider-side costs for spam filtering are comparable 
to an arrangement where the provider classifies plaintext emails
non-privately. %

For the topic extraction module, the cost improvements relative to the
baseline~(\S\ref{s:starting}) are smaller: provider \cpu drops by
$1.37\times$, client \cpu drops by $3.25\times$, storage goes up by a
factor of $2$, and the network cost goes up slightly. Beyond that, the
\emph{non}-private version of this function is vastly cheaper
than for spam, to the point that the private version is up to two orders
of magnitude worse than a non-private version (depending on the
resource). The next section addresses this. %

\subsection{Rearranging and pruning in \interestextraction}
\label{s:protocoluse}
\label{s:pruning}

\paragraph{Decomposed classification.}
So far, many of the costs are proportional to $B$: storage
(see Figure~\ref{f:gllm+yao}, ``setup phase''),
\cpu cost of Yao, and network cost of Yao (step~\ref{e:yao}). 
For spam filtering, this is not a problem ($B=2$) but for
\interestextraction, $B$ can be in the thousands. 

\Sys's response is a technique that we called \emph{decomposed
classification}.
To explain the idea, we regard \interestextraction as abstractly mapping an
email, together with a set $S$ of cardinality $B$ (all possible topics),
down to a set $S^*$ of cardinality 1 (the chosen topic), using a
proprietary model. \Sys decomposes this map into two:

\begin{myenumerate}

\item[(i)] Map the email, together with the set $S$, to a set $S'$ of
cardinality $B'$ (for example, $B'=20$); $S'$ comprises
\emph{candidate topics}. The client does this by itself. %

\item[(ii)] Map the email, together with $S'$, down to a set $S''$ of
cardinality 1; ideally $S''$ is the same as $S^*$ (otherwise, accuracy
is sacrificed). This step relies on a proprietary model and is done
using secure two-party machinery. Thus, the expensive part of the
protocol is now proportional to $B'$ rather than $B$.

\end{myenumerate}

For this arrangement to make sense, several requirements must be met.
First, the client needs to be able to perform the map in step~(i)
locally. Here, \sys exploits an observation: \emph{topic lists (the set
$S$) are public today}~\cite{googleinterests}. They have to be, so that
advertisers can target and users can set interests. Thus, a client can
in principle use some \emph{non}-proprietary classifier for step~(i).
\Sys is agnostic about the source of this classifier; it could be
supplied by the client, the provider, or a third party.

Second, the arrangement needs to be accurate, which it is when
$S'$ contains $S^*$. \Sys observes
that although the classifier used in step~(i) would not be honed, \emph{it
doesn't need to be, because it is performing a far coarser task than
choosing a single topic}. Thus, in principle, the step~(i) map might
reliably produce 
accurate outputs---meaning that the true topic, $S^*$, is among the $B'$
candidates---without much training, expertise, or other proprietary input.
Our experiments confirm that indeed the loss of end-to-end accuracy is
small~(\S\ref{s:topiceval}). 

Finally, step (ii) must not reveal $S'$ to the provider, since that
would be more information than a single extracted topic. This rules out
instantiating step~(ii) by naively applying the existing
protocol~(\S\ref{s:starting}--\S\ref{s:packing}), with $S'$ in
place of $S$. \Sys's response is depicted in
Figure~\ref{f:pruningprotocol}.  There are some low-level details to
handle because of the interaction with packing~(\S\ref{s:packing}), but
at a high level, this protocol works as follows. The provider supplies
the entire proprietary model (with all $B$ topics); the client obtains
$B$ dot products, in encrypted form, via the inexpensive component of
\yaogllmname (secure dot product). The client then extracts and blinds
the $B'$ dot products that correspond to the candidate topics. The
parties finish by using Yao to privately identify the topic that
produced the maximum.

\paragraph{Feature selection.}
Protocol storage is proportional to $N$
(Figure~\ref{f:gllm+yao}, ``setup phase'').
\Sys's response is the standard technique of \emph{feature
selection}~\cite{tang2016toward}: incorporating into the model the
features most helpful for discrimination. This takes place in the
``setup phase'' of the protocol (the number of rows in the provider's
matrix reduces from $N$ to $N'$).  Of course, one presumes
that providers already prune their models; the proposal here is to do so
more aggressively.  Section~\ref{s:topiceval} shows that in return for
large drops in the number of considered features, the accuracy drops
only modestly. In fact, reductions of 75\% in the number of features is
a plausible operating point.

\paragraph{Cost savings.}
    The client-side storage costs reduce by the factor $N/N'$ due
    to feature selection.
    For $B{=}2048, B'{=}20$, and $L{=}692$ (average number of
    features per email in the authors' emails), relative to the protocol in 
    \S\ref{s:packing},
    the provider \cpu drops by $45\times$, client
    \cpu drops by $8.4\times$, and the network transfers drop by $20.4\times$ (\S\ref{s:topiceval}).
Thus, the aforementioned two orders of magnitude (above the non-private
version) becomes roughly 5$\times$.

\subsection{Robustness to misbehaving parties}
\label{s:maliciousparties}

\Sys aims to provide the following guarantees, even when parties deviate
from the protocol:
\begin{myenumerate2}
\item The client and provider cannot (directly) observe each other's inputs
nor any intermediate state in the computation.
\item The client learns at most 1 bit of output each time spam
classification is invoked.
\item The provider learns at most $\log B$ bits of output per email.
This comes from \interestextraction.
\end{myenumerate2}

Guarantee~1 follows from the baseline protocol, 
which includes
mechanisms that thwart the attempted subversion of the
protocol~(\S\ref{s:starting}). Guarantee~2 follows from Guarantee~1
and the fact that the client is the party who gets the spam
classification output. Guarantee~3 follows similarly, provided that the
client feeds each email into the protocol at most once; we discuss this
requirement shortly. 

Before continuing, we note that the two applications are not symmetric.
In spam classification, the client, who gets the output, could conceivably
try to learn the provider's model; however,
the provider does not directly learn anything about the client's email.
With \topicextraction, the roles are reversed.  Because the output is obtained
by the provider, what is potentially at risk is the privacy of the
email of the client, who instead has no access
to the provider's model.

\paragraph{Semantic leakage.} Despite its guarantees about the
\emph{number} of output bits, \sys has nothing to say
about the \emph{meaning} of those bits. For example, in \topicextraction, an
adversarial provider could construct a tailored ``model'' to attack an
email (or the emails of a particular user), in which case the $\log B$
bits could yield important semantic information about the email.  A
client who is concerned about this issue has several options, including
opting out of \topicextraction (and presumably compensating the provider
for service, since a key purpose of \topicextraction is ad display, which
generates revenue). We describe a more mischievous response below.

In the spam application, an adversarial client could construct emails to
try to pinpoint or infer model parameters, and then leak the model to
spammers (and more generally undermine the proprietary nature of the
model). A possible defense would be for the provider to periodically
revise the model (and maintain different versions). 

\paragraph{Repetition and replay.}
An adversarial provider could conceivably replay a given email to
a client $k$ times.
The provider would then get $k \log B$ bits from the email,
rather than $\log B$.
Our defense is simply for the client to regard email transmission
from each sender (and possibly each sender's device) as a separate asynchronous---and 
    lossy and duplicating---transmission channel. Solutions to detecting
    duplicates over such channels are well-understood: counters,
    windows, etc.
Something to note is that, for this defense to work,
 emails have to be signed, otherwise an adversary can
 deny service by pretending to be a sender and spuriously exhausting 
 counters.

\paragraph{Integrity.} \Sys does not offer any guarantees about which
function Yao actually computes. For example, for \interestextraction,
the client could garble~(\S\ref{s:yao}) argmin rather than argmax. In
this case, the aforementioned guarantees continue to hold (no inputs are
disclosed, etc.), though of course this misbehavior interferes with the
ultimate functionality.  \sys does not defend against this case, and in
fact, it could be considered a feature---it gives the client a passive
way to ``opt out'', with plausible deniability (the client could garble
a function that produces an arbitrary choice of index).

The analogous attack, for spam, is for the provider to garble a function
other than threshold comparison. This would undermine the spam/nospam
classification and would presumably be disincentivized by the same
forces incentivizing providers to supply spam filtering as a service in
the first place.

\section{Other details and implementation}
\label{s:impl}
\label{s:implementation}

\paragraph{Keyword search in \sys.}
\label{s:search}
So far \sys's design has focused on modules for classification, 
but for completeness, \sys also needs to handle a third function: keyword
search.
\Sys's variant of keyword search is a simple existence proof that
the provider's servers
are not essential: the client simply maintains and queries a client-side search
index.
This brings a storage cost, but our evaluation suggests that it is
tolerable~(\S\ref{s:searcheval}).
One issue is that an index may not be available if a user logs in from a
new machine. This can be addressed with a provider-side solution,
which is future work (such a solution could be built
on searchable
symmetric encryption, which allows search over encrypted data~\cite{cash13sse, jarecki13outsourced, cash2014dynamic,
pappas14blind}). 

\paragraph{Implementation details.}
We have fully implemented the design described in
\S\ref{s:arch}, \S\ref{s:starting}, \S\ref{s:innovations}, and the
keyword search module just described.
We borrow code
from existing libraries:
    Obliv-C~\cite{zahur2015obliv} for Yao's 2PC protocol;\footnote{%
Another choice would have been
TinyGarble~\cite{songhori2015tinygarble}. We found the performance of Obliv-C
and TinyGarble to be comparable for the functions we compute inside Yao in \sys; we
choose the former because it is easier to integrate with \sys's C{++}
code.}
XPIR~\cite{melchor2016xpir} for the \xpirname AHE
scheme, \textsc{liblinear}~\cite{fan2008liblinear, liblinearimpl} to train LR
and SVM classifiers, and SQLite FTS4~\cite{sqlitefts} for the search index.
Besides the libraries, \sys is written in 5{,}300 lines of C{++} 
and 160 lines of Python.

\section{Evaluation}
\label{s:eval}
\label{s:evaluation}

Our evaluation answers the following questions:
\begin{myenumerate}
    \item What are the provider- and client-side overheads of \sys? For what
    configurations (model size, email size, etc.) are they low? 

    \item How much do \sys's optimizations help in reducing the overheads?

    \item How accurate are \sys's functions (for example, how accurately can \sys
    filter spam emails or extract topics of emails)?
\end{myenumerate}
A summary of evaluation results is as follows:
\begin{myitemize}

\item \Sys's provider-side \cpu consumption for spam filtering and \interestextraction
is, respectively, 0.65 and 1.03--1.78$\times$ of a non-private
arrangement,
and, respectively, 0.17$\times$ and 0.01--0.02$\times$ of its baseline~(\S\ref{s:starting}).
\item Network transfers in \sys are 2.7--5.4$\times$ of a non-private
arrangement,
and 0.024--0.048$\times$ of its baseline~(\S\ref{s:starting}).
\item \Sys's client-side \cpu consumption
is less than 1s per email, and storage space use is a few 
hundred MBs. These are a few factors lower
than in the baseline~(\S\ref{s:starting}).
\item For \interestextraction,
the potential coarsening effects of \sys's classifiers (\S\ref{s:pruning}) are
a drop in 
accuracy of between 1--3\%.

\end{myitemize}

\emparagraph{Method and setup.}
We consider 
spam filtering, \interestextraction, and keyword
search separately.
For spam filtering and \interestextraction, we compare \sys to its
starting baseline (\S\ref{s:starting}) (which we call \textbf{\hybrid}), and
\textbf{\nopriv}, which models the status quo, in which the provider locally runs
classification on plaintext email contents. %
For the keyword search function, we consider only the %
basic client-side search index based scheme
(\S\ref{s:impl}).

We vary the following parameters: number of features ($N$) and
categories ($B$) in the classification models,
number of features in an
email ($L$), and the number of candidate topics ($B'$) in \interestextraction.
For the classification models, we use both synthetic and real-world datasets.
To generate synthetic emails, we use %
random words (between 4 to 12 letters each). For real-world data, 
we use the \lingspam~\cite{androutsopoulos2000evaluationlingspam} ($481$ spam and
$2{,}411$ non-spam emails),
Enron~\cite{enronwebsite} ($17{,}148$ spam and $16{,}555$ non-spam emails of about 150 Enron
employees) and Gmail
($355$ spam and $600$ non-spam emails from the Gmail account of one of the
authors)\footnote{Gmail stores spam for only a limited
time; therefore, we do not have access to a large number of spam emails. The non-spam
emails are from the same time period as the spam emails.}
datasets for spam filtering evaluation, and
the 20 Newsgroup~\cite{20newswebsite} ($18{,}846$ Usenet posts on 20 topics),
Reuters-21578~\cite{reutersdswebsite} ($12{,}603$ newswire stories on 90 topics),
and RCV1~\cite{lewis2004rcv1} ($806{,}778$ newswire stories from 296 regions) 
datasets for \interestextraction evaluation.

We measure resource overheads in terms of 
provider- and
client-side \cpu times to process an email,
network transfers between provider and client, and the storage space used at a
client. 
The resource overheads are independent of the classification algorithm (NB, LR,
SVM),
so we present them once; the accuracies depend on the classification
algorithm, so we present them for each algorithm.

Our testbed is Amazon EC2. We use one %
machine of type m3.2xlarge for the provider
and one machine of the same type for a client.
At the provider, we use an independent \cpu for each
module/function.
Similarly, the client uses a single \cpu.

\begin{figure}[t]
\centering

\newcommand{\cellbreakline}[2][b]{%
  \begin{tabular}[#1]{@{}l@{}}#2\end{tabular}}

\resizebox{\columnwidth}{!}{

\begin{tabular}{l}

    \begin{tabular}{rrrrr}
     & encryption & decryption & addition & left shift and add \\
     \cmidrule(lr){2-2}
     \cmidrule(lr){3-3}
     \cmidrule(lr){4-4}
     \cmidrule(lr){5-5}
    GPG &  \SI{1.7}{\milli\second} & \SI{1.3}{\milli\second} & N/A & N/A\\
    Paillier & \SI{2.5}{\milli\second} & \SI{0.7}{\milli\second} & \SI{7}{\micro\second} & N/A \\
    \xpirname & \SI{103}{\micro\second} & \SI{31}{\micro\second} & \SI{3}{\micro\second} & \SI{70}{\micro\second} \\

    \end{tabular}
    \\

\specialrule{0.6pt}{6pt}{4pt}

    \begin{tabular}{rrr}
    Yao cost & \cpu &
    network transfers \\
     \cmidrule(lr){2-2}
     \cmidrule(lr){3-3}
    $\phi$ = integer comparison & \SI{71}{\micro\second} & $2501$ B\\
    $\phi$ = argmax & \SI{70}{\micro\second} & $3959$ B\\

    \end{tabular}
    \\

\specialrule{0.6pt}{6pt}{4pt}

    \begin{tabular}{>{\hspace{0.7cm}}rrr}
         & map lookup & float addition\\
     \cmidrule(lr){2-2}
     \cmidrule(lr){3-3}
        \nopriv operations & \SI{0.17}{\micro\second} & \SI{0.001}{\micro\second} \\
    \end{tabular}
    \\

\bottomrule

\end{tabular}

}

\normalfont\selectfont
\caption{%
Microbenchmarks for the common operations (Figure~\ref{f:costs}) in \sys
and the baselines. 
\cpu costs come
from a single \cpu
of an Amazon EC2 machine of type m3.2xlarge. Both \cpu and network costs are
averaged over 1{,}000 runs; standard deviations (not shown) are within 5\% of
the averages.
GPG encryption/decryption times depend on the length of the
email; we use an email size of
$75$~KB, which is in
line with average email size~\cite{avgemailsize2}.  
Similarly, Yao costs for $\phi{=}\text{argmax}$ depend linearly on the number of input values; we show
costs per input value.
}
\label{fig:microbenchmarks}
\label{f:microbenchmark}
\label{f:microbenchmarks}
\end{figure}

\emparagraph{Microbenchmarks.}
\label{s:microeval}
\label{s:microbenchmarks}

Figure~\ref{fig:microbenchmarks} shows the 
\cpu and network costs for the common operations (Figure~\ref{f:costs}) 
in \sys and the baselines.
We will use these microbenchmarks to explain
the performance evaluation in the next subsections.

\subsection{Spam filtering}
\label{s:spameval}

This subsection reports the resource overheads (provider- and client-side \cpu
time, network transfers, and client-side storage space use) and accuracy of
spam filtering in \sys.

We set three different values for the number of features in the
spam classification model: $N {=} \{200\textrm{K}, 1\textrm{M}, 5\textrm{M}\}$. These values
correspond to the typical number of
features in various deployments of Bayesian spam filtering
software~\cite{spamdbsize11m, spamdbsize15m, spamdbsize150k}. %
We also vary the number of features in an email ($L {=} \{200, 1000,
5000\}$); these values are chosen based on the \lingspam dataset (average of $377$
and a maximum of $3638$ features per email) and the Gmail dataset (average of $692$
and a maximum of $5215$ features per email).
The number of categories $B$ is two: spam and non-spam.

\begin{figure}[t]
\centering
\includegraphics[width=3.15in]{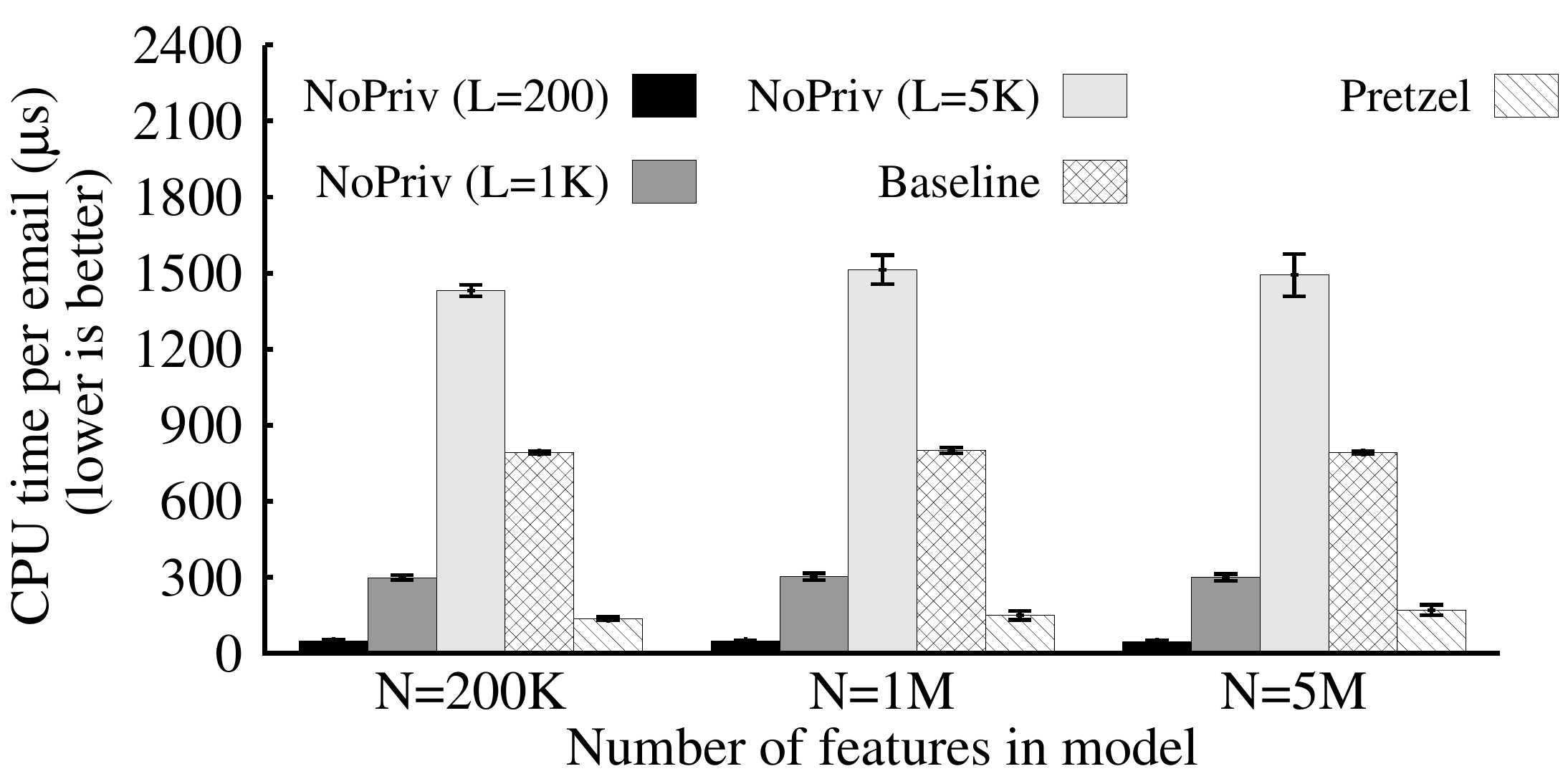}
\caption{Provider-side \cpu time per email in microseconds for the
spam filtering module while varying the number of features ($N$) in the spam
classification model, and the number of features ($L$)
in an email. %
\cpu time for \nopriv varies only slightly with $N$ (not visible),
while (provider-side) \cpu times for \hybrid and \sys
are independent of both $L$ and $N$ (Figure~\ref{f:costs}). 
}
\label{f:spamprovidercputime}
\label{fig:spamprovidercputime}
\end{figure}

\emparagraph{Provider-side \cpu time.}
Figure~\ref{f:spamprovidercputime} shows the per-email \cpu time consumed by the
provider. 

For emails with fewer features ($L{=}200$), the \cpu time of \sys is $2.7\times$ \nopriv's and
$0.17\times$ \hybrid's.
\sys's is more than \nopriv's because in \nopriv the provider does $L$ feature extractions,
map lookups, and float additions, which are fast operations
(Figure~\ref{fig:microbenchmarks}), whereas in \sys, the provider does
relatively expensive operations: one additively homomorphic decryption
of a \xpirname{} ciphertext plus one comparison inside Yao
(Figure~\ref{f:costs} and \S\ref{s:refinement1}). %
\Sys's \cpu time is lower than \hybrid's because 
in \sys, the provider decrypts a \xpirname{} ciphertext whereas in \hybrid the
provider decrypts a Paillier
ciphertext %
(Figure~\ref{f:microbenchmarks}).

As the number of features in an email increases ($L{=}\{1000, 5000\}$),
the provider's \cpu time in both \sys and
\hybrid does
not change, as it is independent of $L$ (unlike
the client's)
while \nopriv's increases
since it is linear in $L$ (see Figure~\ref{f:costs}).
A particular point of interest is 
$L{=}692$ (the average number of features per email 
in the Gmail dataset),
for which \cpu time of \sys is
$0.65\times$ \nopriv's.
\begin{figure}[t]
\footnotesize
\centering

\begin{tabular}{
@{}
*{1}{>{\raggedright\arraybackslash}b{.14\textwidth}}  @{ } %
*{1}{>{\raggedleft\arraybackslash}b{.08\textwidth}}  %
*{1}{>{\raggedleft\arraybackslash}b{.08\textwidth}}  %
*{1}{>{\raggedleft\arraybackslash}b{.09\textwidth}}  %
*{1}{>{\raggedleft\arraybackslash}b{.09\textwidth}}  %
@{}
}

& \multicolumn{3}{c}{Size} \\

\cmidrule(l){2-4}

& $N{=}200$K & $N{=}1$M & $N{=}5$M \\    
    \midrule
    Non-encrypted & 4.3 MB & 21.5 MB & 107.3 MB \\
    \hybrid & 51.6 MB & 258.0 MB & 1.3 GB\\
    \sys-NoOptimPack & 3.1 GB & 15.3 GB & 76.3 GB\\
    \sys & 7.4 MB & 36.7 MB & 183.5 MB\\
    \bottomrule
    
\end{tabular}
\normalfont\selectfont
\caption{%
Size of encrypted and plaintext spam classification models. $N$ is
the number of features in the model. \sys-NoOptimPack is \sys with the
legacy packing technique~(\S\ref{s:packing}).}
\label{fig:spammodelsize}
\label{f:spammodelsize}
\end{figure}

\emparagraph{Client-side overheads.}
Figure~\ref{f:spammodelsize} shows the size of the spam model for the various
systems. %
We notice that the model in \sys is approximately $7\times$ smaller
than
the model in \hybrid. This is due to \sys's ``across row'' packing technique
(\S\ref{s:packing}).
We also notice that, given its other refinements, \sys's packing technique
is essential (the \sys-NoOptimPack row in the figure).

In terms of client-side \cpu time, 
 \sys takes $\approx358$~ms to process an email with many features
 ($L{=}5000$)
against 
a
large model ($N{=}5$M). 
This time is dominated
 by the $L$ left shift and add operations in the secure dot product computation (\S\ref{s:packing}).
Our microbenchmarks (Figure~\ref{fig:microbenchmarks}) explain this number:
$5000$ of the left shift and add operation takes
$5000 \times 70\mu\textrm{s}{=}350\textrm{ms}$.
A large $L$ is an unfavorable scenario for \sys: 
client-side processing is proportional to $L$ (Figure~\ref{f:costs}).

\emparagraph{Network transfers.}
\label{s:spamnetwork}
Both \sys and \hybrid add network overhead relative to \nopriv. 
It is, respectively, 
$19.6$~KB and $3.6$~KB per email (or
$26.1$\% and $4.8$\% of \nopriv, when considering average email size as
reported by~\cite{avgemailsize2}).
These overheads are due to transfer of a ciphertext, and a comparison
inside Yao's framework
(Figure~\ref{f:gllm+yao}).
\sys's overheads are higher than \hybrid's because the \xpirname{} ciphertext in
\sys is much larger than the Paillier ciphertext.
\begin{figure}[t]
\footnotesize
\centering

\begin{tabular}{r@{\hspace{0.1in}}r@{\hspace{0.07in}}r@{\hspace{0.07in}}rr@{\hspace{0.07in}}r@{\hspace{0.07in}}rr@{\hspace{0.07in}}r@{\hspace{0.07in}}r}

& \multicolumn{3}{c}{\lingspam} 
& \multicolumn{3}{c}{Enron} 
& \multicolumn{3}{c}{Gmail} 
\\

\cmidrule(l){2-4}
\cmidrule(l){5-7}
\cmidrule(l){8-10}

& Acc. & Prec. & Rec. & Acc. & Prec. & Rec. & Acc. & Prec. & Rec. \\
    
    \midrule
    \binaryNB & 99.4 & 98.1 & 98.1 & 98.8 & 99.2 & 98.4 & 98.1 & 99.7 & 95.2 \\
    LR & 99.4 & 99.4 & 97.1 & 98.9 & 98.4 & 99.5 & 98.5 & 98.9 & 97.2 \\
    SVM & 99.4 & 99.2 & 97.5 & 98.7 & 98.5 & 99.0 & 98.5 & 98.9 & 97.2 \\
    GR  & 99.3 & 98.1 & 97.9 & 98.8 & 99.2 & 98.4 & 98.1 & 99.7 & 95.2 \\

    \bottomrule
    
\end{tabular}

\normalfont\selectfont
\caption{%
Accuracy (Acc.), precision (Prec.), and recall (Rec.) for spam filtering in \sys. 
Sets of columns correspond to the different spam datasets, and the rows
correspond to the classification algorithms \sys supports: 
\binaryNB, 
binary LR, and two-class SVM (\S\ref{s:classification}).
Also shown is accuracy for Grahams-Robinson's original Naive Bayes
algorithm (GR). 
}
\label{fig:spamaccuracy}
\label{f:spamaccuracy}
\end{figure}

\emparagraph{Accuracy.}
Figure~\ref{f:spamaccuracy} shows \sys's spam classification accuracy for the
different classification algorithms it supports. (The figure also shows
precision and recall. Higher precision means lower false positives, or
non-spam falsely classified as spam; higher recall means lower false
negatives, or spam falsely classified as non-spam.) 

\subsection{Topic extraction}
\label{s:topiceval}

This subsection reports the resource overheads (provider- and client-side \cpu
time, network transfers, and client-side storage space use) and accuracy of
\interestextraction in \sys.

We experiment with $N{=}\{20\textrm{K}, 100$K$\}$)\footnote{The number of features in \interestextraction models
are usually much lower than spam models because of many different word
variations for spam, e.g.,
FREE and FR33, etc.}
and $B = \{128, 512, 2048\}$.
These parameters are based on the total number of features 
in the \interestextraction datasets we use
 and Google's public
list of topics (2208 topics~\cite{googleinterests}).
For the number of candidate topics for \sys (\S\ref{s:pruning}), we experiment
with $B' {=} \{5, 10, 20, 40\}$.

\begin{figure}[t]
\centering
\includegraphics[width=3.15in]{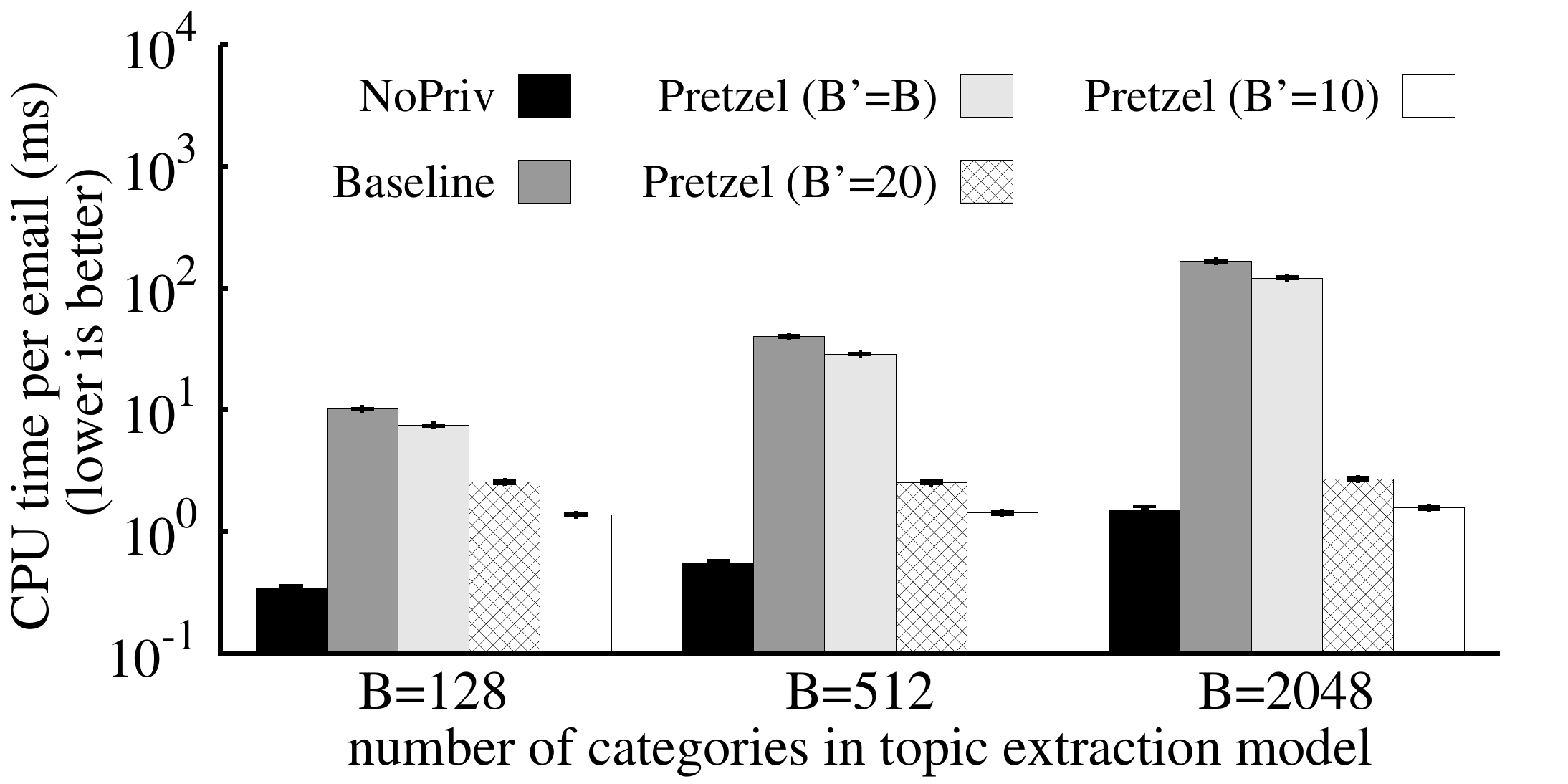}
\caption{%
Provider \cpu time per email in milliseconds for 
\interestextraction,
varying the number of categories ($B$) in the
model and the number of candidate topics ($B'$). 
The case $B=B'$ measures \sys without the decomposed classification
technique~(\S\ref{s:pruning}).
The y-axis is
log-scaled. $N$ and $L$ are set to $100$K and $692$ (average number of features
per email
in the authors' Gmail dataset).
The \cpu times do not depend on $N$ or $L$ for 
\sys and \hybrid; they increase linearly with $L$ and vary slightly with $N$ for \nopriv.}
\label{f:topicprovidercputime}
\label{fig:topicprovidercputime}
\end{figure}

\emparagraph{Provider-side \cpu time}
Figure~\ref{f:topicprovidercputime} shows the per email \cpu time consumed by the
provider.

Without decomposed classification~(\S\ref{s:pruning})---this is the
$B'=B$ case in the figure---\sys's \cpu time is significantly higher
than \nopriv's but lower than \hybrid's.
The difference between \sys and \hybrid reflects two facts.
First, \xpirname{} ciphertexts are
cheaper to decrypt than Paillier ciphertexts. Second,
there are fewer \xpirname{} ciphertexts, because each is larger, and 
can thus pack in more plaintext elements~(\S\ref{s:packing}). 
With decomposed classification,
the number of comparisons inside Yao's framework come down and, as expected,
the difference between \cpu times in \sys and \nopriv drops~(\S\ref{s:pruning}).
For $B{=}2048, B'{=}20$, \sys's \cpu time is $1.78\times$ \nopriv's; for
$B{=}2048, B'{=}10$, it is $1.03\times$ \nopriv's.

\begin{figure}[t]
\footnotesize
\centering

\begin{tabular}{
@{}
*{1}{>{\raggedright\arraybackslash}b{.13\textwidth}}  @{ } %
*{1}{>{\raggedleft\arraybackslash}b{.08\textwidth}}  %
*{1}{>{\raggedleft\arraybackslash}b{.08\textwidth}}  %
*{1}{>{\raggedleft\arraybackslash}b{.09\textwidth}}  %
*{1}{>{\raggedleft\arraybackslash}b{.08\textwidth}}  %
@{}
}

& \multicolumn{3}{c}{network transfers} \\

\cmidrule(l){2-4}

& $B{=}128$ & $B{=}512$ & $B{=}2048$ \\
    \midrule
    \hybrid & 501.5 KB & 2.0 MB & 8.0 MB\\
    \sys ($B'{=}B$) & 516.6 KB & 2.0 MB & 8.0 MB\\
    \sys ($B'{=}20$) & 402.0 KB & 402.0 KB & 401.9 KB\\
    \sys ($B'{=}10$) & 201.0 KB & 201.0 KB & 201.2 KB\\
    \bottomrule

\end{tabular}
\normalfont\selectfont
\caption{%
Network transfers per email for the \interestextraction module in \sys
and \hybrid. $B'$ is the number of candidate topics in decomposed classification
(\S\ref{s:pruning}). Network transfers are independent of the number of
features in the model ($N$) and email ($L$) (Figure~\ref{f:costs}).}
\label{fig:topicnetwork}
\label{f:topicnetwork}
\end{figure}

\emparagraph{Network transfers.}
Figure~\ref{f:topicnetwork} shows the network transfers per email for \hybrid and
\sys. %
As expected, with decomposed classification, \sys's network transfers are lower;
they are 
$402$~KB per email, or $5.4\times$ the average email
size of 
74KB, as reported in~\cite{avgemailsize2}
 for $B{=}2048,
B'{=}20$, and
$201$~KB per email (or $2.7\times$ the average email size) for
$B{=}2048, B'{=}10$.

\begin{figure}[t]
\footnotesize
\centering

\begin{tabular}{
@{}
*{1}{>{\raggedright\arraybackslash}b{.14\textwidth}}  @{ } %
*{1}{>{\raggedleft\arraybackslash}b{.08\textwidth}}  %
*{1}{>{\raggedleft\arraybackslash}b{.08\textwidth}}  %
*{1}{>{\raggedleft\arraybackslash}b{.09\textwidth}}  %
@{}
}

& \multicolumn{2}{c}{Size} \\

\cmidrule(l){2-3}

& $N{=}20$K & $N{=}100$K  \\
    \midrule
    Non-encrypted & 144.3 MB & 769.4 MB \\
    \hybrid & 288.4 MB & 1.5 GB \\
    \sys & 720.7 MB & 3.8 GB \\
    \bottomrule
    
\end{tabular}
\normalfont\selectfont
\caption{%
Size of \interestextraction models for the various systems. $N$ is the number of features in
the model. $B$ is set to $2048$.}
\label{fig:topicmodelsize}
\label{f:topicmodelsize}
\end{figure}

\emparagraph{Client-side overheads.}
Fig.~\ref{f:topicmodelsize} shows the model sizes 
(before feature selection \S\ref{s:pruning})
for the various systems for different values of $N$ and $B{=}2048$.
\sys's model is bigger than \hybrid's for two reasons. First, its model
comprises an encrypted part that comes from the provider and 
a public part. Second, the \xpirname{}
ciphertexts used in the encrypted part
have a larger expansion factor (by a factor of 2) than Paillier
ciphertexts.

In terms of client-side \cpu time, 
as in spam filtering, \sys (with or without pruning) takes less than half a
second to process an email with many features ($L{=}5000$).

\begin{figure}[t]
\centering
\includegraphics[width=3.15in]{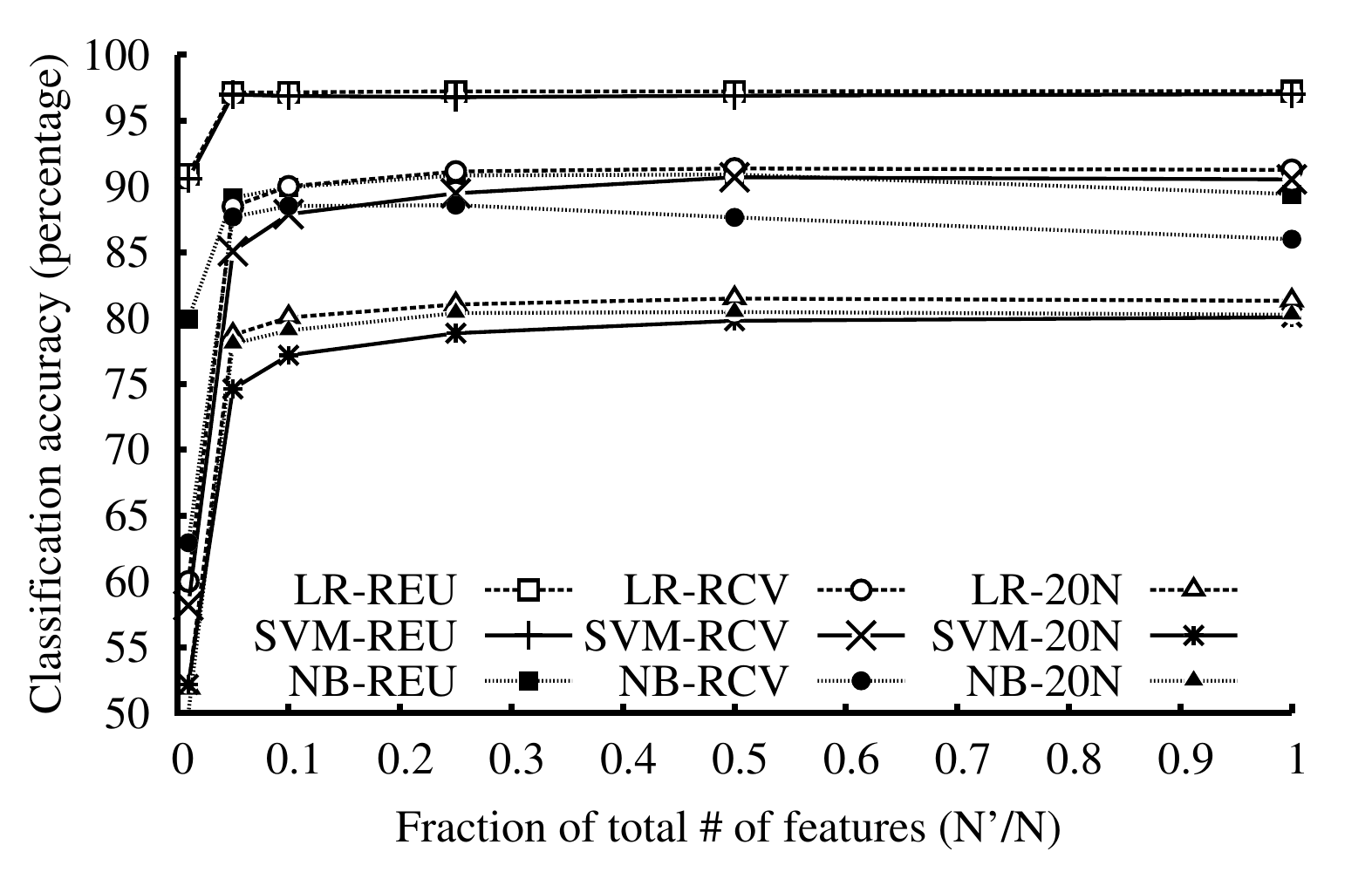}
\caption{%
Classification accuracy 
of \interestextraction classifiers
in \sys
as a function of the degree of feature selection $N'/N$
(\S\ref{s:pruning}). %
($N$ is the total number of features in the training part of the datasets 
and $N'$ are the number of selected features.)
(The 20News and Reuters datasets come pre-split into training and
testing parts: 60\%/40\% and 75\%/25\% for the two respectively; we randomly
split RCV1 into 70\%/30\% training/testing portions.)
\sys can operate at a point where number of features selected
$N'$ are roughly 25\% of $N$; this would result in only a marginal drop in
accuracy.
}
\label{f:topicaccuracy}
\label{fig:topicaccuracy}
\end{figure}

\begin{figure}[t]
\footnotesize
\centering

\begin{tabular}{
@{}
*{1}{>{\raggedright\arraybackslash}b{.08\textwidth}}  @{ } %
*{1}{>{\raggedleft\arraybackslash}b{.06\textwidth}}  %
*{1}{>{\raggedleft\arraybackslash}b{.06\textwidth}}  %
*{1}{>{\raggedleft\arraybackslash}b{.06\textwidth}}  %
*{1}{>{\raggedleft\arraybackslash}b{.06\textwidth}}  %
}

& \multicolumn{4}{c}{Percentage of the total training dataset} \\
\cmidrule(lr){2-5}
            & 1\% & 2\% & 5\% & 10\% \\
\midrule

$B'{=}5$   & 79.6 & 84.0 & 90.1 & 94.0 \\
$B'{=}10$  & 89.6 & 92.1 & 95.6 & 97.7 \\
$B'{=}20$  & 95.9 & 97.3 & 98.5 & 99.3 \\
$B'{=}40$  & 98.7 & 99.3 & 99.8 & 99.9 \\

\bottomrule
\end{tabular}

\normalfont\selectfont
\caption{Impact of decomposed classification (\S\ref{s:pruning}) on classification accuracy for the
RCV1 dataset with 296 topics.
The columns (except the first) correspond to the percentage of the total training dataset
used to train the (public) model that extracts candidate topics. The rows correspond to
the number of candidate topics ($B'$).
The cells contain the percentage of test documents for which the ``true
category'' (according to a classifier trained on the entire
training dataset) is contained in the candidate topics. Higher percentage is better; 100\% is ideal.}
\label{fig:pruning}
\label{f:pruning}
\end{figure}

\emparagraph{Loss of accuracy.}
Recall that classification accuracy for \interestextraction in \sys could be
affected by feature selection and decomposed classification
(\S\ref{s:pruning}). 
Figure~\ref{f:topicaccuracy} shows classification accuracy 
for \topicextraction classifiers trained with and without feature
selection, and while varying the degree of feature selection
(using the Chi-square selection technique~\cite{tang2016toward}).
It appears that even after a high degree of feature selection, accuracy drops
only modestly below its peak point.
(This would reduce the client-side storage cost presented in
Figure~\ref{f:topicmodelsize}.)

Figure~\ref{f:pruning} shows the variation in classification accuracy 
due to decomposed classification~(\S\ref{s:pruning}).
(The depicted data are for the RCV1 dataset and NB classifier; 
the qualitative results are similar for the other datasets and classifiers.)
The data suggest that an effective non-proprietary classifier 
can be trained
using a small fraction of training data, for only a small loss in
end-to-end accuracy.

\subsection{Keyword search and absolute costs}
\label{s:searcheval}

\begin{figure}[t]
\footnotesize
\centering

\begin{tabular}{
@{}
*{1}{>{\raggedright\arraybackslash}b{.17\textwidth}}  @{ } %
*{1}{>{\raggedleft\arraybackslash}b{.08\textwidth}}  %
*{1}{>{\raggedleft\arraybackslash}b{.08\textwidth}}  %
*{1}{>{\raggedleft\arraybackslash}b{.08\textwidth}}  %
@{}
}

& index size & query time & update time \\    
    
    \midrule
    \lingspam & 5.2 MB & 0.32 ms & 0.18 ms \\
    Enron & 27.2 MB & 0.49 ms & 0.1 ms \\
    20 Newsgroup & 23.9 MB & 0.3 ms & 0.12 ms \\
    Reuters-21578 & 6.0 MB & 0.28 ms & 0.06 ms \\
    Gmail Inbox (40K emails) & 50.4 MB & 0.13 ms & 0.12 ms \\
    \bottomrule
    
\end{tabular}
\normalfont\selectfont
\caption{%
Client-side search index sizes, \cpu times to query a keyword in
the indexes (i.e., retrieve a list of emails that contain a keyword),
and \cpu times to index a new
email.}
\label{fig:search}
\label{f:ssearch}
\end{figure}

Figure~\ref{fig:search} shows the client-side storage and \cpu costs 
of \sys's keyword search module (\S\ref{s:impl}).
\label{s:evaldiscussion}
We now consider the preceding costs in absolute terms, with a view
toward whether they would be
acceptable in a deployment.
We consider an average user who receives
$150$ emails daily~\cite{avgemailsperday} of average size
($75$KB)~\cite{avgemailsize2}, and owns a mobile device with
$32$~GB of storage.

To spam filter a long email, the client takes $358$~ms, which would be less than
a minute daily. As for the encrypted model, one with $5$M features occupies
$183.5$~MB or 0.5\% of the device's storage. For network overheads, each
email transfers an extra 19.33~KB, which is $2.8$~MB daily.

For \interestextraction, the client uses less than
half a second of \cpu per email (or less than $75$s daily);
a model with $2048$ categories (close to Google's) and $20$K features
occupies $720.7$MB or $2.2$\% of the device's storage (this can be reduced
further using feature selection); and, a $B'{=}20$ pruning
scheme transfers an extra $59$~MB ($5.4$ times the size of the emails) over the
network daily.

Overall, these costs are certainly substantial---and we don't mean to
diminish that issue---but we believe that the magnitudes in question are
still within tolerance for most users.

\section{Discussion, limitations, and future work}
\label{s:discussion}
\label{s:limitations}
\label{s:futurework}
\label{s:obstacles}

\sys is an improvement over its baseline~(\S\ref{s:starting}) of
up to 100$\times$, depending on the resource~(\S\ref{s:eval}). %
Its
absolute overheads are substantial but, as just
discussed~(\S\ref{s:evaldiscussion}), are
within the realm of plausibility.

\sys's prototype has many limitations. It does not handle services 
beyond what we presented;
extending \sys to other
functions
(predictive personal assistance, virus scanning, etc.) is 
future work. So is extending \sys to hide metadata, perhaps
by composing with anonymity systems and other
secure computation primitives.

A fundamental limitation of \sys is information
leakage~(\S\ref{s:designethos}, \S\ref{s:maliciousparties}).  As
discussed in Section~\ref{s:maliciousparties}, the principal concerns
surround the privacy of providers' models (for spam) and clients' emails
(during \topicextraction). While we don't have technical defenses,
 we note that participation in \topicextraction is
voluntary for the client; in fact, a client can opt out with plausible
deniability~(\S\ref{s:maliciousparties}). Moreover, providers cannot
expose all or even a substantial fraction of clients this way, as that
would forfeit the original purpose of \topicextraction. Nevertheless, we
are aware that, defaults being defaults, most clients would not opt out,
which means that particular clients could indeed be targeted by a
sufficiently motivated adversarial provider. 

If \sys were widely deployed, we would need a way to derive and retrain
models. This is a separate problem, with existing research~\cite{kantarcioglu2003privacy,
yi2009privacy,
sumana2014privacy,
gangrade2012privacy,
zhan2008privacy,
yang2005privacy,
vaidya2004privacy,
gangrade2013privacy,
zhan2005privacy,
vaidya2008privacy,
zhan2004privacy,
vaidya2008privacySVM,
yu2006privacyhorizontal,
yu2006privacyvertical}; an
avenue of future work for \sys would be to incorporate these works.

There
are many other obstacles between the status quo and default end-to-end
encryption. In general, it's hard to modify a communication medium as
entrenched as email~\cite{zimmermannopgp}. On the other hand, there is
reason for hope: TLS between data centers was deployed over just several
years~\cite{durumeric15neither}. Another obstacle is key management and
usability: how do users share
keys across devices and
find each other's keys? This too is difficult, but there is recent research
and commercial attention~\cite{keybase,whatsapp,melara2015coniks}.
Finally, politics: there are entrenched interests who would
prefer email not to be encrypted.

Ultimately, our goal is just to demonstrate an alternative.
We don't claim that \sys is an optimal point
in the three-way
tradeoff among functionality, performance, and
privacy~(\S\ref{s:designethos}); we don't yet know what such an
optimum would be.
We simply claim that it is different from the status quo
(which combines rich functionality,
superb performance, but no encryption by default) and
that it is potentially plausible.

\ifx\buildtechreport\undefined
\else
\appendix
\section{Some details of Naive Bayes}

\subsection{Spam classification}
\label{a:spamderivation}

Section~\ref{s:classification} stated that the \binaryNB algorithm
labels a document (email) as spam if $\alpha=$\mbox{$1/p(\textrm{spam} \given \vec{x}) -
1$} is
greater than some fixed threshold, and that $\log \alpha$
can be expressed as  
\begin{align}
\nonumber &\left(\sum_{i=1}^{i=N} x_i \cdot \log{p(t_i \given C_2)}\right) + 1
\cdot \log {p(C_2)} \\ 
- &\left(\sum_{i=1}^{i=N} x_i \cdot \log {p(t_i \given C_1)}\right) + 1
\cdot \log {p(C_1)},
\label{eq:spamlog}
\end{align}
where the spam category is represented by $C_1$ and non-spam is $C_2$.
Here we show this derivation.

From Bayes Rule:
\begin{equation*}
    p(C_1 \given \vec{x}) = \frac{p(\vec{x} \given C_1)\cdot
    p(C_1)}{p(\vec{x})}.
\end{equation*}
Rewriting:
\begin{align}
\nonumber \frac{1}{p(C_1 \given \vec{x})} - 1 &= \frac{p(\vec{x})}{p(\vec{x} \given C_1) \cdot
p(C_1)} - 1. \\
\intertext{Expanding $p(\vec{x})$ to
$p(\vec{x} \given C_1) \cdot p(C_1) + p(\vec{x} \given C_2) \cdot p(C_2)$:}
&= \frac{p(\vec{x} \given C_2) \cdot p(C_2) }{p(\vec{x} \given C_1) \cdot p(C_1)}.
\label{eq:eqfoo}
\end{align}
The Naive Bayes assumption says that the 
    presence (or absence) of a feature is independent of
    the other features, which means
\begin{equation*}
\label{eq:spamfirststep}
     p(\vec{x} \given C_j) = \prod_{i=1}^{i=N} p(t_i \given C_j) ^ {x_i}.
\end{equation*}
Plugging the above equation into equation~\eqref{eq:eqfoo}:
\begin{align*}
\alpha =  \frac{\prod_{i=1}^{i=N} p(t_i \given C_2) ^ {x_i} \cdot p(C_2)}{\prod_{i=1}^{i=N}
p(t_i \given C_1) ^ {x_i} \cdot p(C_1)}.
\end{align*}
In log space, the expression above is 
equation~\eqref{eq:spamlog}.

\subsection{Multinomial text classification}
\label{a:multinomialderivation}

Section~\ref{s:classification} stated that the multinomial NB
algorithm
    identifies the category
    $C_{j^*}$ that maximizes $j^* = \text{argmax}_j\, p(C_j
    \given \vec{x})$.
The section stated that 
    the maximal $p(C_j \given \vec{x})$ can be determined by examining
\begin{align*}
&\left(\sum_{i=1}^{i=N} x_i \cdot \log{p(t_i \given C_j)}\right) + 1
\cdot \log {p(C_j)},
\end{align*}
for each $j$. Here we give a derivation of this fact.

From Bayes Rule:
\begin{equation}
\label{eq:bayesrule}
    p(C_j \given \vec{x}) = \frac{p(\vec{x} \given C_j)\cdot
    p(C_j)}{p(\vec{x})}.
\end{equation}
The multinomial Naive Bayes algorithm assumes that a document, email, or feature
vector $\vec{x}$ is formed by picking, with replacement, $L{=}\sum_{i=1}^{i=N} x_i$
features from the set of $N$ features in the model.
Then, $p(\vec{x} \given C_j)$ is modeled as a multinomial distribution:
\begin{equation*}
     p(\vec{x} \given C_j) = \frac{p(L) \cdot L!}{\prod_{i=1}^{i=N} x_i!} \cdot
\prod_{i=1}^{i=N} p(t_i \given C_j)^{x_i}. %
\label{eq:multinomialfirststep}
\end{equation*}
Plugging the above equation into Equation~\eqref{eq:bayesrule}:
\begin{align*}
     &= \frac{1}{p(\vec{x})} \cdot \frac{p(L) \cdot L!}{\prod_{i=1}^{i=N} x_i!} \cdot
\prod_{i=1}^{i=N} p(t_i \given C_j)^{x_i} \cdot p(C_j)
\end{align*}
Discounting the terms that are independent of $j$ (which we are free to
do because the maximum is unaffected by such terms),
the expression above becomes:
\begin{align*}
\prod_{i=1}^{i=N} p(t_i \given C_j)^{x_i} \cdot p(C_j), 
\end{align*}
which, when converted to log space, is the claimed expression.

\fi
\frenchspacing

{
\goodcitationsize
\begin{flushleft}
\setlength{\parskip}{0pt}
\setlength{\itemsep}{0pt}
\bibliographystyle{abbrv}
\bibliography{conferences-long-with-abbr2,paper}
\end{flushleft}
}
\label{p:last}
\end{document}